\documentclass[final,3p,times,authoryear]{elsarticle}

\usepackage{graphicx}
\usepackage{verbatim}
\usepackage{amsmath}
\usepackage{amssymb}
\usepackage{amsfonts}
\usepackage{mathrsfs}
\usepackage{adjustbox}
\usepackage{mdframed}
\usepackage{eurosym}
\DeclareRobustCommand{\officialeuro}{%
  \ifmmode\expandafter\text\fi
  {\fontencoding{U}\fontfamily{eurosym}\selectfont e}}
\usepackage{lipsum}

\usepackage{dutchcal}

\graphicspath{ {figures/} }

\usepackage{nomencl}
\usepackage{etoolbox}
\usepackage{wrapfig}
\usepackage{framed} 
\usepackage[skins,breakable, most]{tcolorbox}

\renewcommand*\nompreamble{\begin{multicols}{2}}
\usepackage{epsfig}
\usepackage{epstopdf} 
\usepackage{todonotes}
\usepackage{setspace}
\usepackage{siunitx}

\usepackage[utf8x]{inputenc} 
\usepackage{stfloats}

\usepackage[caption = false]{subfig}
\usepackage[hyphens]{url}
\usepackage{graphicx}
\usepackage{booktabs}

\makeatletter
\newcommand{\myonecolumn}{\clearpage \global \columnwidth \textwidth \global \hsize \columnwidth \global \linewidth \columnwidth \global \@twocolumnfalse \col@number \@ne \@floatplacement}
\makeatother

\usepackage{lmodern,textcomp}
\usepackage{rotating} 
\usepackage{multicol} 
\usepackage{multirow} 
\usepackage{mathtools, cuted}
\usepackage{lipsum}
\usepackage{tabularx} 
\newcolumntype{L}[1]{>{\raggedright\arraybackslash}p{#1}} 
\newcolumntype{C}[1]{>{\centering\arraybackslash}p{#1}} 
\newcolumntype{R}[1]{>{\raggedleft\arraybackslash}p{#1}} 
\usepackage{amssymb}
\usepackage{pifont}
\RequirePackage[ngerman=ngerman-x-latest]{hyphsubst}
\usepackage{microtype}
\usepackage[american]{babel} 
\usepackage{hyperref}
\usepackage{float}
\usepackage{algorithm}
\usepackage{algorithmicx}
\usepackage{algpseudocode}

\usepackage{amsmath}

\usepackage{lineno}

\biboptions{square,comma,sort&compress}

\setcitestyle{numbers}
\journal{Applied Energy}

\renewcommand{\nompreamble}{This list presents relevant abbreviations and symbols that are used within the body of this work.}

\renewcommand\nomgroup[1]{%
  \item[\bfseries
  \ifstrequal{#1}{A}{Abbreviations}{%
  \ifstrequal{#1}{I}{Indices}{%
  \ifstrequal{#1}{P}{Parameters}{%
  \ifstrequal{#1}{V}{Variables}{%
  \ifstrequal{#1}{M}{Vectors \& Matrices}{%
  \ifstrequal{#1}{G}{Graphs}{%
  \ifstrequal{#1}{Y}{Systems}{%
  \ifstrequal{#1}{C}{Scalars}{}}}}}}}}%
]}
\makeglossary
\makenomenclature

\usepackage{tikz}
\usetikzlibrary{arrows,shapes,snakes,automata,backgrounds,petri, positioning}
\usetikzlibrary{matrix,calc}

\begin{document}

\title{From passive to active: flexibility from electric vehicles in the context of transmission system development}

\author[DTU1]{Philipp Andreas Gunkel\corref{cor1}}
\ead{phgu@dtu.dk}
\author[DTU1]{Claire Bergaentzl\'e}
\author[DTU2]{Ida Græsted Jensen}
\author[DTU2]{Fabian Scheller}
\cortext[cor1]{Corresponding author}

\address[DTU1]{Energy Economics and Regulation, DTU Management, Technical University of Denmark, Kongens Lyngby, Denmark}
\address[DTU2]{Energy Systems Analysis, DTU Management, Technical University of Denmark, Kongens Lyngby, Denmark}
\begin{abstract}
Electrification of transport in RES-based power system will support the decarbonisation of the transport sector. However, due to the increase in energy demand and the large peak effects of charging, the passive integration of electric cars is likely to undermine sustainability efforts.
This study investigates three different charging strategies for electric vehicle in Europe offering various degrees of flexibility: passive charging, smart charging and vehicle-to-grid, and puts this flexibility in perspective with the flexibility offered by interconnections.
We use the Balmorel optimization tool to represent the short-term dispatch and long-term investment in the energy system and we contribute to the state-of-the-art in developing new methodologies to represent home charging and battery degradation.
Our results show how each step of increased charging flexibility reduces system costs, affects energy mix, impacts spot prices and reduces CO2 emissions until the horizon 2050. We quantify how flexible charging and variable generation mutually support each other (>100TWh from wind and solar energy in 2050) and restrict the business case for stationary batteries, whereas passive charging results in a substitution of wind by solar energy. The comparison of each charging scheme with and without interconnection expansion highlights the interplay between European countries in terms of electricity prices and CO2 emissions in the context of electrified transport. Although the best outcome is reached under the most flexible scenario at the EU level, the situation of the countries with the cheapest and most decarbonised electricity mix is damaged, which calls for adapted coordination policy at the EU level. 
\end{abstract}

\begin{keyword}
Renewable energy, Electric vehicle, Transmission system, Energy system, Flexibility
\end{keyword}

\maketitle

\section*{Highlights}
\begin{itemize}
\item Modelling impacts of electric vehicle flexibility compared to grid expansion
\item Electric vehicle flexibility substitutes solar power with wind power
\item Marginal effect on stationary battery deployment through grid expansion
\item Local reduction of electricity price fluctuations by electric vehicle flexibility
\item Electricity price fluctuations are spatially smoothed by grid expansion
\end{itemize}

\section{Introduction}
Europe has ambitious plans to decarbonize the energy system, including in sectors like transport. As a consequence, sector coupling is a challenge that has to be addressed while switching generating technologies in the direction of variable renewable energies (VRE). Inevitably, vehicles running on fossil fuels are replaced by electric vehicles (EVs), increasing the need for production units in the electricity sector. These new production units must then cover the new demand at the required times, which influences investment decisions. However, the specific demand requirements of EVs are highly dependent on the chosen charging schemes.  

A typical EV more than doubles an average household’s electricity consumption \cite{yilmaz2012review}. According to Friedl et al. \cite{friedl2018blackout} as well as Kasten et al. \cite{kasten2016assessing}, to various extents distribution grids are not sufficient to serve the demand for the expected deployment of EVs. Friedl et al. \cite{friedl2018blackout} state that, with an EV share of 30\%, supply bottlenecks will occur if one applies the current state-of-the-art charging scheme. Passive charging (PC) can be seen as a threat to the successful integration of EVs, since it puts great pressure on existing electricity grids and require large peak power capacities. In this context, two solutions are conceivable \cite{friedl2018blackout}: extending the grid, which requires long periods of preparation; or avoiding passive charging behavior, which might even generate economic benefits. Kasten et al. \cite{kasten2016assessing} also suggest the introduction of dispersed control of the charging process to deal with technical restrictions in the grid infrastructure. Smart-charging (SC) options as a way of shifting demand could create various benefits for both the mobility sector and the electricity grid. The PC process that begins when a car is connected to the grid and stops when the battery is full could be replaced by a controlled smart-charging process that shifts the demand to low load hours. For example, nighttime charging is shown to have a minimal impact on the power grid when coupled with sufficient automation and an adequate control system \cite{yilmaz2012review}. Also, demand shifting does not cause additional battery wear \cite{wang2016quantifying} and can smooth out peak loads and reduce charging costs, assuming consumer prices reflect the wholesale market \cite{uddin2018viability}.
Moreover, Kempton and Tomić \cite{kempton2005vehicle} show that on average, cars are on the road only around 4\% of the time. Thus, EV batteries could potentially be used for other purposes the remaining 96\% of the time, for example, for vehicle-to-grid (V2G) applications. The term “vehicle-to-grid” refers to “the reciprocal flow of power between an electric vehicle and a recipient” \cite{uddin2018viability}. V2G stands for a bidirectional coordinated energy flow going from the grid to the EV unit, namely charging, but also for a coordinated energy flow going from the EV unit to the grid, namely discharging. This transforms the EV unit into flexible storage where the electricity can be used for trading on the spot market or to provide not only negative, but also positive reserve power. \\
In contrast, the term “smart charging” is used only for the unidirectional process.
Previous research from Gan et al. \cite{gan2012optimal} or Sortomme and El-Sharkawi \cite{sortomme2010optimal} showed that even unidirectional smart charging can significantly help to smooth energy demand and stabilize the system. However, bidirectional V2G seems to be even more promising. Sortomme and El-Sharkawi \cite{sortomme2011optimal} found that the increased flexibility of the power supply and the potentially reduced peak load can help integrate renewable energy sources into the system while still providing additional profit. It has also been shown that such findings do not face impossible challenges in terms of the optimization problem that a potential aggregator would be confronted with \cite{han2010development}. In this context, Guille and Gross \cite{guille2009conceptual} elaborate on frameworks for implementing V2G technologies from both a technical and a practical point of view, with the theoretical goal of delaying the use of cycling and peak load units by several hours in any given day. \\
To determine the additional burden of EVs, Hanemann et al. \cite{hanemann2017grid} give an overview of worst-case peak demand in previous studies. They also presents a sophisticated model that maps a multitude of combustion engine driving patterns in a German traffic study into corresponding, distinctive EV equivalents \cite{hanemann2017grid}. The generated load curves show that the peak demand by EVs is higher on work days than at weekends. Furthermore, pure EVs should be distinguished from plug-in hybrid electric vehicles (PHEVs). The generated load curves have also been utilized for optimization studies to determine the effect of charging options as PC, SC and V2G \cite{hanemann2017effects} and further developed in \cite{hanemann2018effects}. A similar analytical approach has been pursued by Wu et al. \cite{wu2010electric}, and a simulation model has been developed by Harris and Webber \cite{harris2014empirically}. Both studies use the American National Household Travel Survey.\\
Different authors have also integrated EVs and PHEVs into energy-system models in order to assess the market effects of static and flexible demand respectively. {{\O}}stergaard et al. \cite{stergaard2015EnergyDemand} predicted developments in the Danish energy system in 2050 with inflexible and flexible EV and heat pumps using the EnergyPLAN model and found a significant impact on energy systems. The charging patterns forced changes in the use of wind and condensing power capabilities. More wind energy was used to cover EV demand when flexible charging was enabled, which, however, just shifted the demand to later hours. A quite similar study of charging behavior during the night and/or daytime for the Danish power market has been conducted by Kristoffersen et al. \cite{kristoffersen2011optimal}. Juul and Meibom \cite{Juul2012Road2030} investigated the electrification of northern Europe’s transport sector. A transport add-on in the Balmorel energy system model allowed for investment choices in different vehicle types. Three different charging options are also implemented: PC, SC and V2G. In this context, Graabak et al. \cite{graabak2016optimal} demonstrate only minor impacts of EVs on the Nordic power system. Whereas Juul and Meibom \cite{Juul2012Road2030} have shown that EVs can reduce fluctuations in electricity prices. Moreover, VRE production has become more important. Also, the total costs were lower the more flexible the charging schemes were, and emissions declined. The study referred to the period up to 2030 but resulted in large computational times. Reduced total costs with smart charging compared to dumb charging have also been determined by Madzharov et al. \cite{madzharov2014integrating} as well as Kiviluoma and Meibom \cite{kiviluoma2011methodology}. Major reasons are seen in the shifting of demand of EVs to hours characterized by low operating costs by replacing thermal plant production. Other energy-system models work through renewable generators and by reducing peak-load power plants. Models like PyPSA \cite{Brown2018PyPSA:Analysis} also allow the integration of EVs in a simplified manner as shown in the research papers of Brown et al. \cite{Brown2018BatteryCharging}. This methodology considers home and work charging and results in reduced system costs, as well as the smoothing effect of variable renewable production \cite{Brown2018SynergiesSystem}. However, the study assumes a large availability of vehicles, as cars are always plugged into a charger when they are not used. Also, the effect of degradation was not taken into account, which can be particularly challenging given the enforced state of charge (SOC) goals of between 75-100\%. EV scheduling can be largely affected by battery ageing costs, resulting in stronger load shifting towards the morning hours and spreading charging hours with less power \cite{Gunkel2018OptimizedUncertainties}. Overall, the V2G charging strategy has been proven to be most beneficial 
\cite{hanemann2018effects}. At the same time, different aspects that are relevant for the successful deployment of V2G applications, such as the economic costs of additional charging cycles causing battery degradation, environmental performance, user behavior and consumer acceptance, remain untouched in the vast majority of assessments \cite{sovacool2018neglected}.\\
The literature also compares the contributions of different sources of flexibility. The impact of tax and tariff schemes in enhancing the flexibility of demand response technologies against the impact of transmission expansion is compared in the research report of Flex4RES \cite{Flex4RES2019FlexibleReport}, Gunkel et al. \cite{Gunkel2020ModellingStudy} focus on approaches to the methodology of transmission systems and compares them to the simplified but smart operation of heat pumps and EVs. Both studies showed synergies between flexibility options and reduced overall system costs. In contrast, Brown et al. \cite{Brown2018SynergiesSystem} investigated flexibility options in demand-side management using simplified assumptions for transmission. European countries were either entirely isolated or connected with infinite transmission capacity. Assessing the impact on the variability of prices and the competition and synergy effects due to flexible EVs and their interplay with the transmission system have therefore not been investigated in depth so far. 
In summary, the techno-economic performance of EVs has been investigated on the basis of different assumptions and various applications. Nevertheless, there is a lack of knowledge about the effect of interplay of flexibility options like transmission system expansion, stationary batteries and sector-coupling. Moreover, previous studies using large scale energy system models overestimated the availability of EV and further did not account for cost from battery degradation, which plays a major role in the scheduling of charging processes. \\
In this context, this study aims to investigate how the provision of additional flexibility through EVs compared to the flexibility provided transmission system expansion affects the long-term development of the European energy system. 
On the one hand, this study compares EV charging schemes and transmission expansion to showcase how the system reacts differently to both types of flexibility. On the other hand, synergy and competition effects between both flexibility options are investigated in order to highlight and quantify their impact on energy system development. Furthermore, the impact on other flexibility options such as stationary batteries and sector-coupling to the heat sector.\\
This is done by a model-based assessment of the technical, economic and environmental impacts in northwest Europe, that is, Central West Europe (CWE), the Nordic and Baltic countries and the UK. The Balmorel energy system model, which combines the short-term operation of electricity and the district-heating sector with long-term capacity investments \cite{Ravn2001TheBackground}, is used to estimate the gains of flexible charging schemes like smart charging and V2G being introduced to the system in different decades from 2020 to 2050. This article fills the gap in existing research in three different ways. First it proposes a new methodology to model both battery electric vehicles and plug-in hybrid vehicles in order to better reflect the respective flexibility characteristics of both vehicles. Pathways for the market uptake of EVs are subsequently optimized. Second, additional cost functions are introduced to the modelling of the calendrical and cyclical degradation of batteries to provide a finer representation of load shifting and the smoothing out of charging hours. Third, a new generic data set is developed to integrate vehicles as aggregated virtual storage with dynamic potentials and requirements in order to include EV demand and flexibility potential efficiently. The benefits of the different charging schemes are estimated in relation to the reductions in system costs, their impact on wholesale price variability and distribution, VRE integration and CO2 emissions. In contrast to previous literature, the option of endogenous transmission expansion compared to and with EV charging schemes show the impact and outreach of market coupling as well as local flexibility in order to highlight substitution effects, synergies and dependencies.\\
The study is structured as follows. Section \ref{sec:EVInt} introduces the charging schemes and the data generation, including vehicle stock projections and driving pattern analysis. Section \ref{sec:Balm} demonstrates integration of the EV add-on into Balmorel, while \ref{sec:Res} gives the results of the examined scenarios. This is followed by a discussion in section \ref{sec:Disc} and conclusions in \ref{sec:conc}.
\section{Flexibility potentials of electric vehicles} \label{sec:EVInt}

\subsection{Flexibility provision through charging schemes}
Charging schemes are at the interface of sector coupling between EV and the electricity system. They determine when and how much energy is needed. Thus, the applied charging scheme has an effect on the operation of the available generating assets, as well as on future investment decisions, to cover demand as and when necessary. In theory, in the long term the flexible charging behavior of EVs will affect energy systems positively. In this study, the impacts of three different charging schemes are subsequently investigated.\\ 
The first scheme is current state-of-the-art passive charging. Intuitively, EV charging is not controlled but starts with full charger capacity when the vehicle is plugged in and stops when it is fully charged. Using this scheme does not permit flexible behavior, and furthermore it usually increases the pressure on the electricity system because the main hours of charging are in the afternoon and evening, when prices are already at their highest. The next charging scheme investigated is smart charging. This is considered to be an option that can provide flexibility to the electricity system. Energy purchases are scheduled on the basis of electricity prices as well as degradation costs. The last option examined here for sector coupling with greater flexibility using EVs is V2G. As with smart charging, vehicles react entirely to market signals and their cost function. However, due to the additional installation of a bidirectional charger, cars can actively discharge their batteries as well. Consequently, EVs can contribute to the energy system by balancing supply and demand in both directions. Therefore, this scheme is seen as the option for delivering the greatest degree of flexibility for the system, though it is also linked to greater costs due to differences in charger technology and the additional costs of battery usage.

\subsection{Consumption and flexibility projection}
\subsubsection{Electric vehicle stocks} \label{sec:VehStockNord}
In order to project the future demand and flexibility potential of EVs, the penetration of the technology must be investigated. Vehicle stock projections for the countries involved are summarized in Tables \ref{tab:VehStockNord} and \ref{tab:VehStockCentr}

\begin{table}[H]

\caption{Assumptions for electric vehicle stock developments of the Nordics and Baltics in thousand vehicles.}
\centering
\label{tab:VehStockNord}
\begin{tabular}{c|r|ccccccc}\hline \hline
Year & Type & DK    & NO     & SE     & FI    & EE    & LV    & LT    \\\hline
2020 & BEV          & 5.0   & 21.0   & 20.0   & 3.0   & 1.4   & 0.5   & 0.5   \\
2020 & PHEV         & 11.0  & 11.0   & 75.0   & 18.0  & 0.5   & 0.5   & 0.5   \\
2030 & BEV          & 169.8 & 1068.2 & 352.5  & 183.3 & 43.5  & 61.7  & 88.1  \\
2030 & PHEV         & 540.4 & 646.9  & 1121.9 & 583.4 & 138.3 & 196.4 & 280.4 \\
2040 & BEV          & 583.1 & 1703.6 & 1182.9 & 615.1 & 156.6 & 222.2 & 317.4 \\
2040 & PHEV         & 832.1 & 1112.7 & 1688.0 & 877.8 & 223.4 & 317.1 & 452.9 \\
2050 & BEV          & 698.3 & 2354.8 & 1502.7 & 781.4 & 202.0 & 242.0 & 314.5 \\
2050 & PHEV         & 761.0 & 1059.4 & 1637.6 & 851.6 & 220.1 & 263.7 & 342.8
\end{tabular}
\end{table}

\begin{table}[H]
\caption{Assumptions for electric vehicle stock developments of central north west Europe in thousand vehicles.}
\centering
\label{tab:VehStockCentr}
\begin{tabular}{c|r|cccccc}\hline \hline
Year & Vehicle & UK      & DE      & NL     & PL     & BE     & FR      \\\hline
2020 & BEV          & 70.0    & 100.0   & 51.0   & 1.5    & 48.0   & 70.9    \\
2020 & PHEV         & 170.0   & 105.0   & 100.0  & 1.5    & 12.5   & 172.2   \\
2030 & BEV          & 3371.1  & 4340.7  & 2872.7 & 1700.0 & 516.3  & 3414.9  \\
2030 & PHEV         & 3371.1  & 4340.7  & 1739.6 & 1700.0 & 516.3  & 3414.9  \\
2040 & BEV          & 7271.1  & 8472.7  & 4183.5 & 3400.0 & 1032.7 & 7365.6  \\
2040 & PHEV         & 5453.3  & 6354.6  & 2732.6 & 2550.0 & 774.5  & 5524.2  \\
2050 & BEV          & 11700.0 & 12396.1 & 5320.7 & 5100.0 & 1549.0 & 11852.1 \\
2050 & PHEV         & 7800.0  & 8264.1  & 2393.8 & 3400.0 & 1032.7 & 7901.4 
\end{tabular}
\end{table}

The vehicle stocks are projected using several sources, values until 2020 being linear extrapolations from \cite{EAFO18}. Country specific projections are taken from \cite{EADTU18} (DK) \cite{TOI16} (NO),  \cite{NER16} (SE, FI, FR, BE),\cite{Summerton2014TheNetherlands} (NL),\cite{Clu15} (UK),\cite{TRAMP06} (DE),\cite{Korolec18} (PL). Where projections do not include shares between BEVs and PHEVs, the shares in \cite{EADTU18} are assumed. Thus, the resulting demand from EVs can be estimated.

\subsubsection{Electric vehicle driving demand}\label{sec:datainp}
The demand from driving is generated using technical assumptions for vehicles and driving data from Denmark’s National Transport Survey \cite{ChristiansenTransportvaneundersgelsenTU0617v2}. In order to quantify the basic energy requirements, technical and economic data regarding both vehicles and chargers are summarized in Table \ref{tab:Flexdat}. The table includes vehicle efficiency $\eta^{Bat,Ch}$, battery storage size of BEVs $\overline{SOC}^{BEV}$, investment cost $C^{Bat}$, charger size $\overline{P}^{Ch}$ and cost $C^{Ch}$ assumptions.   
\begin{table}[ht]
\caption{Yearly dependent technical input data of electric vehicles and charger for the EV-addon.\cite{EADTU18}:*,\cite{Curry2017Lithium-ionMarket}\cite{Lambert2017AudiCars}:**,\cite{Gould2013ATransfer}\cite{2015iv}\cite{Autobild2017StartSchnellladenetz}\cite{Wallboxstore.ch2017InvestmentCharger}:*** }
\label{tab:Flexdat}
\centering
\begin{tabular}{l|ccccccc}\hline \hline
Year  & $\eta^{Veh}$* & $\overline{SOC}^{BEV}$* & $\overline{SOC}^{PHEV}$*  & $\overline{P}^{Ch}$* & $C^{Bat}$** & $C^{Ch}$***\\
$[-]$ & $[-]$ & [$\frac{MWh}{1000 veh}$] & [$\frac{MWh}{1000 veh}$] & [$kW$] & [$\frac{\EUR}{MWh}$] & [$\frac{\EUR}{kWh}$]\\\hline
2020 & 0.18 & 30 & 10 & 0.01  & 175000 & 220.0\\
2030 & 0.17 & 30 & 10 & 0.01    & 140000 & 60.1\\
2040 & 0.16  & 40 & 10 & 0.015    & 105000 & 59.7\\
2050 & 0.15 & 50 & 10 & 0.02     & 70000 & 57.5  
\end{tabular}
\end{table}

The maximum battery size of PHEVs $\overline{SOC}^{PHEV}$* stays constant at 10 $kWh$.
Further technical inputs related to battery degradation are added. Since EV owners have an interest in using their batteries for as long as possible, smarter charging schemes will account for the ageing of the cells. Gunkel \cite{Gunkel2018OptimizedUncertainties} summarizes the most important degradation factors, on the basis of which it also determines cyclical degradation and calendrical ageing to be the relevant factors. Cyclical degradation is dependent on the depth of discharge (DOD), which determines how much of the battery capacity is charged in one cycle. By using the adjusted value suggested by Thingvad and Marinelli \cite{Thingvad2018InfluenceCountries} as well as K{\"o}ll et al. \cite{Koll2011FahrzeugmessungenAZE}, the cells are assumed to age 0.003\% for a full equivalent cycle represented by $\alpha^{Deg,Cyc}$. To reduce computational time, the non-linear expression is simplified to a linear dependency. This simplification was also used for calendrical ageing, which ranges from 0.00006\% to 0.00015\% for every single hour depending on the state of charge of the battery. $\alpha^{Deg,CalC}$ represents the fixed aging of the cells and $\alpha^{Deg,CalF}$ the variable dependent on the SOC. As temperatures are currently disregarded, ageing is the same throughout the year and corresponds to average Danish conditions extracted from empirical data \cite{Thingvad2018InfluenceCountries}. Another factor is used to adjust the battery costs to the general assumption, namely that the lifetime is reached when only 75\% of the original storage capacity remains. This is represented by $\alpha^{Bat,Os}$.\\

\begin{table}[ht]
\caption{Constant technical input data of electric vehicles for the EV-addon.\cite{KommunernesLandsforening2015SaAkutsygehus}:*,\cite{Habib2018AVehicles} :**, \cite{Thingvad2018InfluenceCountries} \cite{Koll2011FahrzeugmessungenAZE} \cite{Gunkel2018OptimizedUncertainties}:***, \cite{Renault-Nissan2017Renault-NissanWarranty}:**** }
\label{tab:Inflexdat}
\centering
\begin{tabular}{cccccccc}\hline \hline
${d}^{BEV}$* & ${d}^{PHEV}$ & $\eta^{Ch,Dch}$**& $\alpha^{Bat,Os}$ ***,**** \\

[$km$] & [$km$] & [-] & [-] \\\hline
50 & 25    & 0.85  & 1.1                \\\hline\hline
$\alpha^{Bat,Lft}$ & $\gamma^{Deg,Cyc}$*** & $\gamma^{Deg,CalC}$ ***& $\gamma^{Deg,CalF}$ ***\\

[-] & [-]& [-] & [-]\\\hline
 0.25 & 0.00004    & 0.0000006    & 0.0000009
\end{tabular}

\end{table}
The generation of reliable driving patterns for vehicle fleets is a key input into the model. On the one hand it determines the electricity demand of the private transport sector. On the other hand, a thorough analysis reveals the potential of EVs to support the energy system and effectively reduce the charging costs. \\
The basis for the generation of driving patterns is interview data from the Danish National Travel Survey (TU) \cite{ChristiansenTransportvaneundersgelsenTU0617v2}. This includes detailed information about vehicle usage, arrival and departure to and from home, and travelled distances. Trips consisting of multiple days are disregarded in this study, and distances refer only to the ranges commuted by the vehicle itself. However, when a vehicle is parked outside home (e.g. at a train station), it is assumed to be unavailable until it returns. Every trip in the TU dataset is further assumed to be equally likely. They are also isolated with regard to the departure and arrival times, as well as distance. Consequently, travel times and distances are not reallocated. This study is limited to home charging due to the limitation of the available dataset. It was not possible to build a statistically strong case for availability pattern representing charging options at workplaces. Potential charging at work and during trips must be considered in future studies, as infrastructure development plans and changes in behavior must be investigated further.\\

To create a data set for regional and national fleets in order to estimate flexibility accurately, every single vehicle is modelled over a full year by itself. Driving behavior is distinguished by weekday, and vehicles are picked randomly from the full set of trips. Two different sets of input data are generated for BEVs and PHEVs respectively, as well as for passive charging and the alternative suitable for smart and V2G charging respectively. The general assumption is that vehicles must have an SOC of 100\% an hour before leaving. The actual hour is determined for every single vehicle based on its earliest departure in the full year for every single weekday. While in the case of PC charging the vehicle is immediately charged when it returns with the maximum charger capacity, the data for the smart charging schemes have a component,  $P^{Inflex}$, which is also an inflexible charging demand when the vehicle has an SOC lower than the minimum emergency requirement, which is based on the minimum range a vehicle should always have. ${d}^{BEV}$ and ${d}^{PHEV}$ represent the minimum distance that the respective car type must satisfy. The residual demand can be freely shifted. Furthermore the charger capacity $\eta^{Ch,Dch}$ stays constant over the entire modelling horizon.\\

\begin{figure}[H] 
    \centering
    \includegraphics[width=0.8\columnwidth]{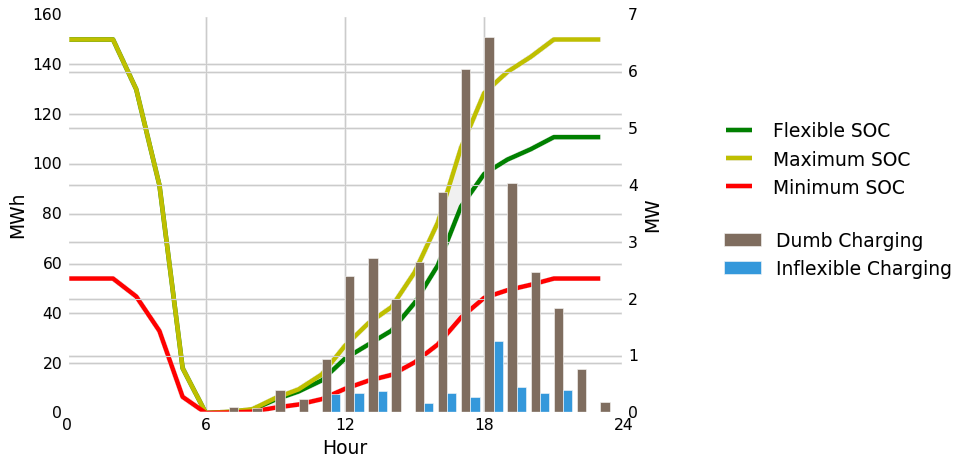} 
     
     \caption{Concept of aggregated virtual storage systems with dynamic requirements for NO1 (Oslo area) in a day in 2020.} 
    \label{fig:VirtStor} 
\end{figure}

As Balmorel only accepts deterministic inputs and due to the complexity of the model, computational time needs to be taken into account. The challenge is therefore to reduce the number of variables to a minimum while at the same time including enough information to describe the flexibility options and the restrictions of every single vehicle accurately. Consequently, an aggregated virtual storage facility with dynamic storage constraints is constructed. Flexible as well as inflexible requirements are maintained by modelling first on the single vehicle level, and then by the aggregation. Figure \ref{fig:VirtStor} visualizes the developed concept with a representative day in NO1 (Oslo area) in 2020.
The yellow line represents the maximum storage capacity of the EV fleet, and the red line shows the minimum capacity. At the beginning of the day, the flexibility of the storage is at its maximum. The actual SOC of the fleet can move freely within those boundaries. In the morning, the fleet must have an SOC of 100\% to cover trips. When vehicles leave, they will reduce the limits corresponding to their battery restrictions and energy levels. Conversely, arriving EVs raise the storage limits again after the trip they have just taken. Since the vast majority of cars do not use their full battery capacity, they will return with more energy than the minimum storage capacity. The virtual SOC is represented by the green line, and the model can freely decide to discharge downwards or charge upwards within the limits. However, some vehicles will arrive with a very low SOC. In order to simulate the immediately required charging, the dark green bars show inflexible and forced consumption. Without this step, the aggregated battery would simply balance the low SOC of some returning EVs with the available flexibility of other cars internally. Consequently, modelling single vehicle patterns is necessary to represent actual flexibility and charging requirements more accurately, while still allowing for aggregation in the latter. The grey bars in the figure show theoretically passive charging behavior as a comparison.\\ 

In order to create the demand for driving, an algorithm is developed to generate the driving pattern of single vehicles and subsequently to allocate all the relevant information into single inputs for the entire fleet. The most important steps are summarized in Alghorithm \ref{alg:VehiclePattern} in the appendix \ref{sec;alg}.

In the first step, a random trip $\omega_{d,c} (t^{Dep},t^{Arr},l^{Dist})$ is chosen out of the sorted weekday-dependent subset $\Omega (d)$ of the overall set of all driving patterns $\Omega$. It contains the departure time $t^{Dep}$, arrival time $t^{Arr}$ and vehicle’s travel distance $l^{Dist}$. The departure and arrival times are rounded to integers corresponding to the hourly time steps in Balmorel. When the vehicle leaves at 09:59, the hour from 09:00 is entirely blocked. When the vehicle returns at 16:00 the entire hour is also blocked, which implies $\beta^{avail}$=0 and that charging can only begin at 17:00. When the vehicle returns home, it is assumed to be plugged in and available ($\beta^{avail}$=1). When the vehicle leaves in the morning, the battery should be fully charged to SOC. It is forced by the model, with trip $\overline{SOC}^{v}$. It is forced by the model with the trip $Tr_{d,c,t}^{Flex}$ representing the charged vehicle battery leaving virtual storage. $P_{d,c,t}^{Inflex}$  represents the inflexible charging demand after the trip if the vehicle comes back with less energy in the battery ($SOC^{Em}$) than is needed to get to the nearest hospital. If the vehicle has more energy than needed in the battery, the state of charge with flexible demand is determined by $SOC_{d,c,t}^{Flex}$. The separation of these data is needed when summing all vehicles to one virtual storage to avoid inner energy balancing. For the passive charging option, $P_{d,c,t}^{Passiv}$ is also calculated after the return. If the charger capacity is less than the required demand, the charged energy is automatically distributed over the subsequent hours using the maximum charger capacity. Afterwards, the minimum and maximum SOCs, $\overline{SOC}_{d,c,t}$ and $\underline{SOC}_{d,c,t}$ respectively, are calculated for every single vehicle based on availability. In the end, all vehicle-dependent values, indicated by the index $c$, are summed over the set $C$ to combine all cars into one large fleet. This is done for battery electric vehicles and plug-in electric vehicles separately. \

Because of the simplification and aggregation from single vehicles to an aggregated fleet, not all the information could be retained. Charger capacity can be overestimated by the model in cases where single vehicles come back with close to full SOC and are charged to the maximum capacity immediately. In the following hour the charger capacity is still there, though due to the aggregated form there is no way of knowing that the vehicle with the specific charger cannot be used anymore. This effect was also studied in a dumb-charging arrangement before application, and it was found that the error is sufficiently minor, forming a normal distribution with maximum values between +/-4\% of the charger capacity while more then 80\% was below +/- 1\%. Therefore, it is assumed to be applicable. \\

\section{Modelling approach}\label{sec:Balm}

\subsection{Balmorel}
This study uses Balmorel, an open-source deterministic energy-system optimization model with a special focus on the electricity and heat sector \cite{Wiese2018BalmorelModel}, \cite{Ravn2001TheBackground}. 
Various add-ons exist to extend the available technologies, fuels and investment decisions. In addition, the transmission system is represented using flow-based modelling. This study covers the Nordic, Baltic and central north-west European countries. Several publications have been published investigating various topics in energy-system modelling, e.g. the impact of CO2-cost on biogas usage \cite{Jensen2017TheUsage}, the future role of district heating in Denmark \cite{Munster2012TheSystem}, the role of waste in the future energy system \cite{Munster2011OptimizationSystem}, and the role of renewable gas in the transition of the electricity and district heating systems \cite{Jensen2020}.
The calibration of the optimization and general modelling assumptions come from the NETP16 and Flex4RES projects \cite{NER16}, \cite{NordicEnergyResearch2019Flex4RESHomepage}. The EV add-on has three different sets of constraints. The first is the set of passive charging (PC) constraints, which is the base case. The second set of constraints relates to the integration of the vehicle with a smart charging-scheme. Constraints for balancing or frequency of market participation are not included since Balmorel only simulates the next-day market. The last set of constraints pertains to the V2G charging scheme.

\subsection{Mathematical formulation of the electric vehicle add-on}
\subsubsection{Passive Charging}

At first, the endogenously determinable PC charging demand $\Omega_{y,s,t}$ [MWh] constraint is presented in equation (\ref{eq:PerDumbObj}). In this equation, $P_{y,a,v,s,t}^{PC}$ [MW] represents the inflexible PC load, dependent on the year $y$, area $a$ in every country, vehicle technology, namely PHEV and BEV $v$, the season $s$ and hour $t$. $\Delta t$ is the length of a time step in the applied optimization model. In this study, Balmorel works with hourly values.

\begin{align}
\label{eq:PerDumbObj}
\Omega_{y,s,t} = \sum_{a}^A \sum_{v}^V P_{y,a,v,s,t}^{PC}\cdot\Delta t
\end{align}

The cost related to the cyclical degradation of PC $\Phi_{y,s,t}^{Deg,Cyc}$ [\EUR] is determined according to equation (\ref{eq:DegCostCycDumb}). For simplicity, it only accounts for charging of the vehicle. The cycle factor $\gamma^{Deg,Cyc}$ represents the cycle degradation of the battery cells. The DOD is determined through the cycled energy per hour $P_{y,a,v,s,t}^{PC}\cdot\Delta t$ divided by the share of the overall installed battery capacity, which is the maximum usable state of charge $\overline{SOC}_{y,a,v,s,t}$ [MWh] of the currently available fleet, and a factor $\alpha^{Bat,Os}$ which increases the installed battery size to account for the difference between usable and installed capacity. Afterwards, the battery replacement cost $C_{y,a,v,s,t}^{Bat,Repl}$ [\EUR] is adjusted by the lifetime factor $\alpha^{Bat,Lft}$, since the end of the battery’s life is assumed after a capacity loss of a quarter. The utilized values of the scalars and parameters are outlined in Table \ref{tab:Inflexdat}.  

\begin{align}
\label{eq:DegCostCycDumb}
\Phi_{y,s,t}^{Deg,Cyc} =  \sum_{a}^A \sum_{v}^V \gamma^{Deg,Cyc}\cdot \frac{C_{y,a,v,s,t}^{Bat, Repl}}{\alpha^{Bat,Lft}}\cdot\frac{P_{y,a,v,s,t}^{PC}\cdot\Delta t}{\alpha^{Bat,Os}\cdot\overline{SOC}_{y,a,v,s,t}}
\end{align}

Degradation costs through calendar ageing of PC $\Phi_{y,s,t}^{Deg,Cal}$ [\EUR] is determined in equation (\ref{eq:DegCostCalDumb}). Since the SOC during trips is not known, the cost is only applied to the battery when it is plugged in. The unavoidable part of ageing is represented in $\gamma^{Deg,CalC}$ while the flexible part, depending on the state of charge, is tied to $\gamma^{Deg,CalF}$ as explained in section \ref{sec:datainp}. The parameter $SOC_{y,a,v,s,t}^{PC}$ [MWh] shows the state-of-charge of the fleet when PC charging is applied.

\begin{align}
\label{eq:DegCostCalDumb}
\Phi_{y,s,t}^{Deg,Cal} &= \sum_{a}^A \sum_{v}^V (\gamma^{Deg,CalC}+\gamma^{Deg,CalF}\cdot\frac{SOC_{y,a,v,s,t}^{PC}}{\alpha^{Bat,Os}\cdot\overline{SOC}_{y,a,v,s,t}})\cdot\frac{C_{y,a,v,s,t}^{Bat,Repl}}{\alpha^{Bat,Lft}}
\end{align}
\subsubsection{Smart Charging}
The next charging scheme to be described is smart charging. The SC charging demand $\Omega_{y,s,t}$ [MWh] must be updated according to equation (\ref{eq:PerG2VObj}). $P_{y,a,v,s,t}^{Inflex}$ [MW] is the inflexible charging which occurs when the vehicle comes back from its trip and contains less energy than the required minimum amount for emergency trips, while $VP_{y,a,v,s,t}^{Flex}$ [MW] is the variable charging load which can be shifted.

\begin{align}
\label{eq:PerG2VObj}
\Omega_{y,s,t} = \sum_{a}^A \sum_{v}^V P_{y,a,v,s,t}^{Inflex}\cdot\Delta t + VP_{y,a,v,s,t}^{Flex}\cdot\Delta t
\end{align}

The cycle degradation costa $\Phi_{y,s,t}^{Deg,Cyc}$ [€] are adjusted for SC in equation (\ref{eq:DegCostCycSmart}) and the calendrical aging cost $\Phi_{y,s,t}^{Deg,Cal}$ [€] in equation (\ref{eq:DegCostCalSmart}). Cyclical costs are now dependent on the sum of the inflexible and flexible charging loads, whereas the calendar ageing cost can now also be reduced by controlling the variable $VSOC_{y,a,v,s,t}$ [MWh], which represents the state of charge of the fleet for SC.

\begin{align}
\label{eq:DegCostCycSmart}
\Phi_{y,s,t}^{Deg,Cyc} =  \sum_{a}^A \sum_{v}^V \gamma^{Deg,Cyc}\cdot \frac{C_{y,a,v,s,t}^{Bat, Repl}}{\alpha^{Bat,Lft}}\cdot\frac{P_{y,a,v,s,t}^{Inflex}+VP_{y,a,v,s,t}^{Flex}}{\alpha^{Bat,Os}\cdot\overline{SOC}_{y,a,v,s,t}}
\end{align}

\begin{align}
\label{eq:DegCostCalSmart}
\Phi_{y,s,t}^{Deg,Cal} &=  \sum_{a}^A \sum_{v}^V (\gamma^{Deg,CalC}+\gamma^{Deg,CalF}\frac{VSOC_{y,a,v,s,t}}{\alpha^{Bat,Os}\cdot\overline{SOC}_{y,a,v,s,t}})\cdot\frac{C_{y,a,v,s,t}^{Bat,Repl}}{\alpha^{Bat,Lft}}
\end{align}

Additional storage and charging constraints must be implemented. Equation (\ref{eq:SOCSmart}) represents the energy balance of the storage, including flexible and inflexible charging, which has efficiency losses $\eta^{Ch,Dch}$ and energy withdrawals through vehicle trips $Tr_{y,a,v,s,t}$ [MWh]. The state of charge $VSOC_{y,a,v,s,t}$ [MWh] has a time-dependent lower and upper constraint $\underline{SOC}_{y,a,v,s,t}$ [MWh] and $\overline{SOC}_{y,a,v,s,t}$ [MWh], which represent the available usable capacity range.
\begin{align}
\label{eq:SOCSmart}
VSOC_{y,a,v,s,t} = &VSOC_{y,a,v,s,t-1} + SOC_{d,c,t}^{Flex} + \eta^{Ch,Dch} (P_{y,a,v,s,t}^{Inflex} + VP_{y,a,v,s,t}^{Flex})\cdot\Delta t - Tr_{y,a,v,s,t} 
\end{align}

\begin{align}
\label{eq:SOCR}
\underline{SOC}_{y,a,v,s,t} \leq VSOC_{y,a,v,s,t} \leq  \overline{SOC}_{y,a,v,s,t}
\end{align}

The operation of the charger is dependent on the maximum power of the vehicle charger $\overline{P}_{y,v}^{Ch}$ [MW] and the number of plugged-in vehicles $Qty_{c,s,t}^{Bat,Avail}$. The sum of flexible and inflexible charging should not exceed the available installed charger capacity, as shown in equation (\ref{eq:CHSMART}), while $VP_{y,a,v,s,t}^{Flex}$ [MWh] cannot be negative as stated in equation (\ref{eq:CHZEROSMART}).

\begin{align}
\label{eq:CHSMART}
0 \leq P_{y,a,v,s,t}^{Inflex} + VP_{y,a,v,s,t}^{Flex} \leq \overline{P}_{y,v}^{Ch}\cdot Qty_{c,s,t}^{Bat,Avail}
\end{align}

\begin{align}
\label{eq:CHZEROSMART}
0 \leq VP_{y,a,v,s,t}^{Flex} 
\end{align}
\subsubsection{Vehicle-to-Grid}
While the SC scheme only allows load to be shifted, the V2G charging scheme also offers to sell energy actively when prices are high. This has a positive effect on the demand constraint for $\Omega_{s,y,t}$ [MWh] as shown in equation (\ref{eq:PerV2GObj}). The option to discharge is represented by the variable $VP_{y,a,v,s,t}^{V2G}$ [MWh] and lowers the cost for EV users as well.
\begin{align}
\label{eq:PerV2GObj}
\Omega_{s,y,t} = \sum_{a}^A \sum_{v}^V P_{y,a,v,s,t}^{Inflex}\cdot\Delta t + VP_{y,a,v,s,t}^{Flex}\cdot\Delta t - VP_{y,a,v,s,t}^{V2G}\cdot\Delta t
\end{align}
The degradation constraints of equations (\ref{eq:DegCostCycSmart}) and (\ref{eq:DegCostCalSmart}) can be both be reused since cyclical degradation is only imposed to charging and the calendar aging is still dependent on the state of charge. The energy balance constraint \ref{eq:SOCSmart} however needs to be updated. Equation (\ref{eq:SOCV2G}) now contains the energy withdrawal of the vehicle from discharging times the efficiency losses.

\begin{align}
\label{eq:SOCV2G}
VSOC_{y,a,v,s,t} =& VSOC_{y,a,v,s,t-1} + SOC_{d,c,t}^{Flex} + \eta^{Ch,Dch}\cdot (P_{y,a,v,s,t}^{Inflex}+VP_{y,a,v,s,t}^{Flex})\cdot\Delta t\\
&- \frac{1}{\eta^{Ch,Dch}}\cdot VP_{y,a,v,s,t}^{V2G}\cdot\Delta t - Tr_{y,a,v,s,t} 
\end{align}

Also the charger constraint \ref{eq:CHSMART} is refreshed accordingly in equation \ref{eq:CHV2G}. The assumption is that it is still not possible to charge and discharge the same vehicle at the same time, but that an entire fleet has the ability to do so. However, the capacity of the virtual charger still limits the degree of charging and discharging corresponding to the installed capacity. It is also a non-negative constraint applied $VP_{y,a,v,s,t}^{V2G}$ in equation (\ref{eq:CHZEROV2G}).

\begin{equation}
\label{eq:CHV2G}
0 \leq P_{y,a,v,s,t}^{Inflex} + VP_{y,a,v,s,t}^{Flex} + VP_{y,a,v,s,t}^{V2G} \leq \overline{P}_{y,v}^{Ch}\cdot Qty_{c,s,t}^{Bat,Avail}
\end{equation}
\begin{equation}
\label{eq:CHZEROV2G}
0 \leq VP_{y,a,v,s,t}^{V2G} 
\end{equation}

\begin{figure}[H] 
    \centering
    \includegraphics[width=0.8\columnwidth]{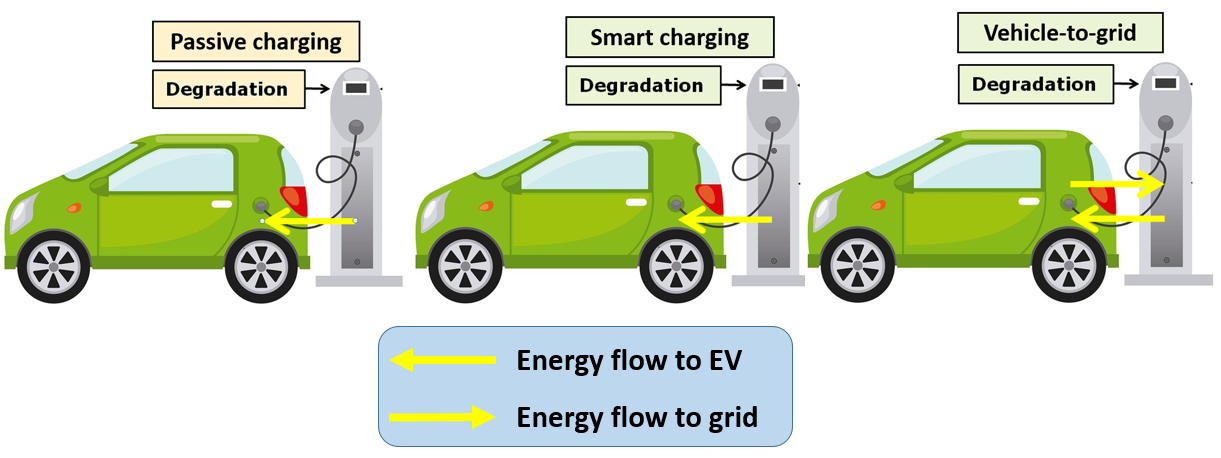} 
    \caption{Indication of energy flows (yellow arrows) for respective charging schemes. Red boxes indicate no optimization of charging and degradation cost and green boxes optimized scheduling of energy in combination with battery degradation cost.} 
    \label{fig:energyflwos} 
\end{figure}

All in all, three types of constraints summarize the influence on energy flows and the interaction with the energy system and EV. Charger constraints limiting energy flows between EV and the grid, storage constraints keeping the stored energy within the physical constraints of the battery and battery degradation cost dependent on the intensity of energy flows as well as the amount of stored energy within the battery.
The corresponding energy flow for passive charging, smart charging and V2G are summarized in figure \ref{fig:energyflwos}. 
The implemented constraints forces energy flows in the passive charging scheme using the full charger capacity until the battery is full. The smart charging constraints optimize energy purchases while taking into account cost from degradation. V2G also optimizes the charging process, however, further allows energy flows back to the grid when price signals of the system indicate profitable hours.\
The constraints for the charging schemes are now introduced and applied in Balmorel. The next section introduces scenarios using the EV add-on.
\section{Scenarios and results} \label{sec:Res}
\subsection{Scenario description}
The scenarios are designed to reflect different strategies of EV integration into the energy system. Table \ref{fig:ScenarioOverview} lists the six simulations.

\begin{table}[h]
\caption{Overview of investigated scenarios and used abbreviation} 
\label{fig:ScenarioOverview} 
\centering
\begin{tabular}{l|cc}\hline \hline
Scenario & Charging scheme & Transmission investments \\ \hline
$PC_{noTransEx}$  & Passive         & Off                      \\
$SC_{noTransEx}$  & Smart           & Off                      \\
$V2G_{noTransEx}$  & V2G             & Off                      \\
$PC_{TransEx}$  & Passive         & On                       \\
$SC_{TransEx}$  & Smart           & On                       \\
$V2G_{TransEx}$  & V2G             & On                      
\end{tabular}
\end{table}

The model is run four times separately from 2020 to 2050, taking over the investment decisions from the previous simulation period. The previously introduced passive, smart and V2G charging schemes are applied in two different cases. The first case investigates the impact of EV flexibility with the transmission system expansion plan until 2030 according to the TYNDP by ENTSO-E, indicated by the abbreviation $_{noTransEx}$. No additional grid expansion is possible in these scenarios after 2030 enabling investigation of the base case and the differences generated by switching the charging schemes. In the second case, all EV charging strategies are run while allowing further grid expansion after 2030. In this case, the electricity system is optimized and provides flexibility in balancing the variable energy resources across markets. These scenarios are indicated by $\_TransEx$ and provide a comparison of the degree of flexibility served by the electric batteries and the transmission system and interconnectors.
\subsection{Scenario outcomes}
\subsubsection{System cost}
\begin{figure*}[ht!]
	\centering
    \subfloat[Total system cost of the scenarios without transmission expansion with $PC_{noTransEx}$ as the base scenario.]{\includegraphics[width=0.8\columnwidth]			           {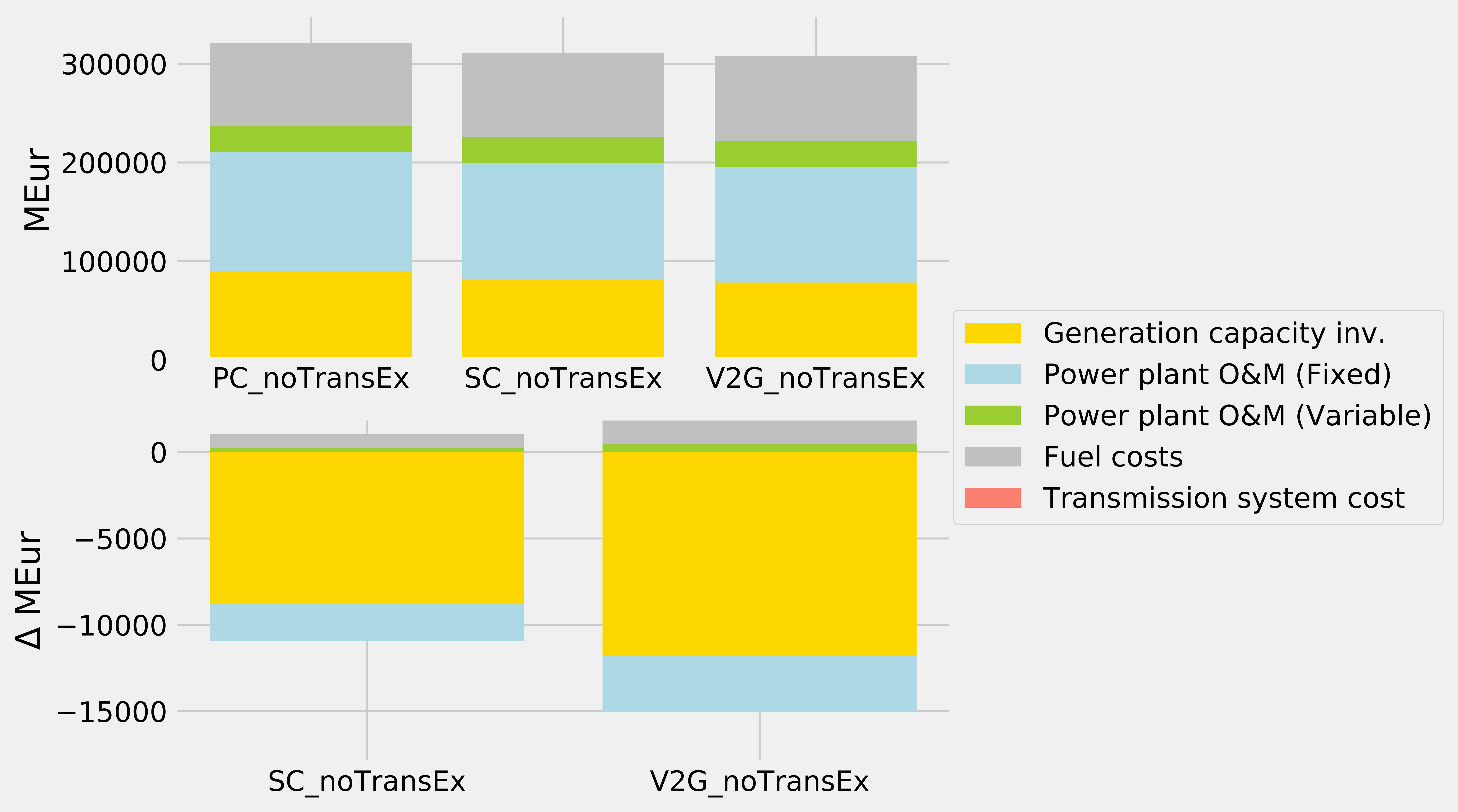}}\\
    \subfloat[Total system cost of the scenarios with transmission expansion with $PC_{TransEx}$ as the base scenario. ]{\includegraphics[width=0.8\columnwidth]
   	{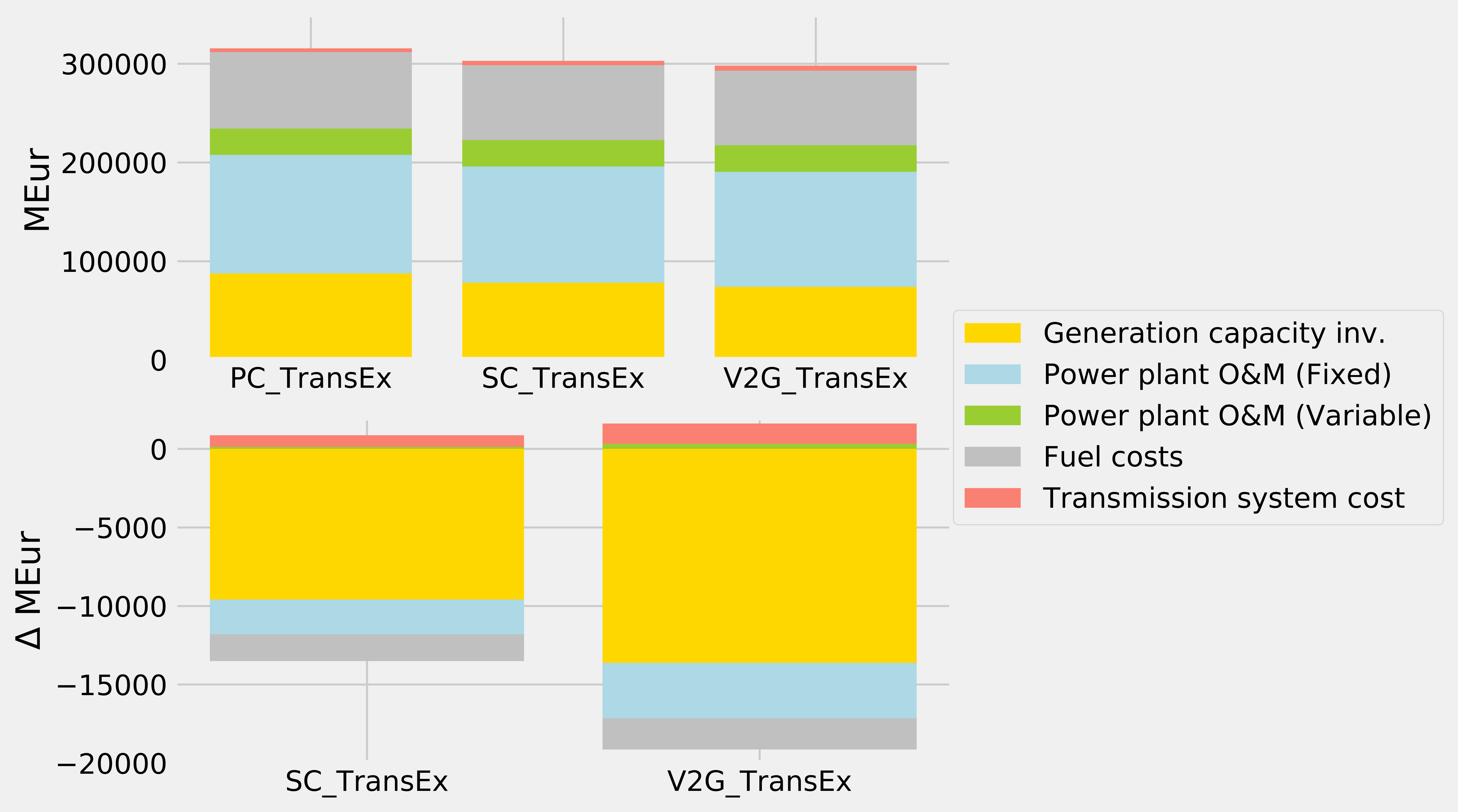}} \\
\centering
\caption{Total system cost of the model. The upper figure illustrates first the cumulated system cost over the entire modelling period while the two bars below show the differences of passive charging compared to smart charging and V2G without transmission expansion. The lower figure shows the total system cost and difference for the charging schemes including transmission line expansion.}
\label{fig:ScenTotCost}
\captionsetup[subfigure]{labelformat=empty}    
\end{figure*}

In order to assess the benefits of the flexible integration of EVs, the total annualized costs of the whole energy system are analyzed. The system cost combines electricity and heat generating plants with network investment costs, including stationary batteries, fixed and variable O\&M and fuel costs. \ref{fig:ScenTotCost} summarizes the cumulated impact of the flexible use of EVs in the SC and V2G scenarios compared to PC relating to the total system cost.
The comparison of the flexibility scenarios shows decreasing system costs with a higher degree of flexibility provided by EV. $PC_{noTransEx}$,the base case, yields an aggregated system costs of 321,091 $M\EUR$ while smart charging, $SC_{noTransEx}$, shows a reduction to system cost of 9,919 $M\EUR$, which corresponds to savings of approximately 3.09\%, and the $V2G_{noTransEx}$ yields a reduction of 4.03\%. The cutting back of capacity investment is the largest contributor to cost savings, especially from other, more expensive storage solutions such as stationary batteries and peak power plants such as gas turbines, thereby showing the substitution role played by EV batteries for balancing. Expenditure on all technologies shrinks except for wind. Fixed O\&M costs have a similar development, since the installed capacity alone is the cause of the payments. Together they amount to overall savings of -11,250 $M\EUR$ and 14,063 $M\EUR$.\\
The only two expenditure items to increase compared to the base case are the fuel cost and the variable costs. Altogether, they cumulate an over-cost of 779 $M\EUR$ and 1,672 M$\EUR$ in the SC and V2G scenario respectively. This is explained by the intensified use of biomass and biogas in the heating sector through cogeneration. In the district heating sector, the flexibility of EVs competes with power-to-heat solutions and leads to higher average prices. Thus, heat pumps are replaced with other technologies, resulting in higher fuel costs because heat pumps can no longer benefit as much from long periods of low prices. Smart charging and V2G, conversely, are charging mainly during these hours. Especially in later decades, the large potential capacity of EV storage pushes out Power-to-Heat (P2H) appliances by charging during previously low price periods in particular. Furthermore, nuclear power is used more and  is more flexible. Conversely, condensing power plants that burn gas are needed less with smart charging and V2G since they shave the peaks, meaning that baseload can produce slightly more energy than before.\\
The lower Figure \ref{fig:ScenTotCost} illustrates the total cumulated system cost and differences with transmission expansion using the three charging schemes. $PC_{TransEx}$ yields 315,544 $M\EUR$, $SC_{TransEx}$ and $V2G_{TransEx}$ 302,905 $M\EUR$ and 298,023 $M\EUR$ respectively. Consequently, when contrasting $V2G_{noTransEx}$ with $V2G_{TransEx}$, additional 3.29 \% can be saved when allowing for line expansions and full EV flexibility. Additional costs from transmission system development ranged between 3,722 and 5,016 $M\EUR$ depending on the scenario. Changing charging schemes, including transmission expansion, also has positive effects on investment expenditure and fixed costs. Compared to $PC_{TransEx}$, smart charging saves 11,797 $M\EUR$ and V2G 17,149 $M\EUR$. Moreover, variable and fuel costs are also reduced between the charging schemes, which was not the case without transmission expansion. $SC_{TransEx}$ has around 1,584 $M\EUR$ and $V2G_{TransEx}$ has 1,667 $M\EUR$ less cost in this segment compared to $PC_{TransEx}$. On the other hand, transmission expansion cost are an additional expenditure. The differences between passive charging to smart charging and V2G are however not large yielding around 741 and 1,672 $M\EUR$. Consequently, the total savings utilizing the full flexibility of transmission expansion and smart charging in $SC_{TransEx}$ amounts to 12,640 $M\EUR$ in savings compared to $PC_{TransEx}$, while $V2G_{TransEx}$ achieves even improvements of as much as 17,522 $M\EUR$.

\subsubsection{Additional flexibility through electric vehicles and market coupling of stationary batteries}
\begin{figure}[H] 
    \centering
    \includegraphics[width=0.8\columnwidth]{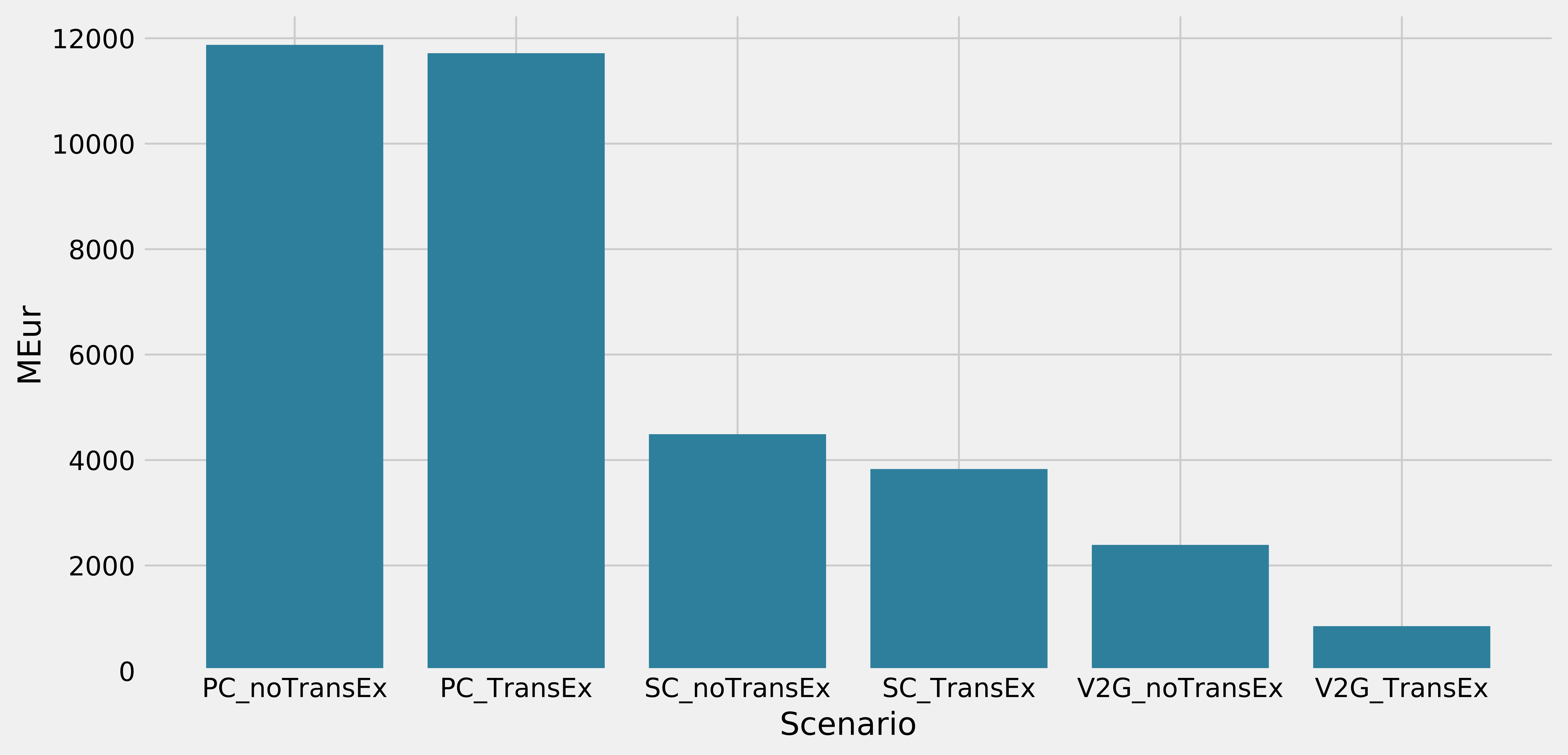} 
    \caption{Accumulated investments into stationary batteries in the entire model space including all years.} 
    \label{fig:Instbat} 
\end{figure}

The load shifting and storage capacities of electric cars have large impacts on competing technologies. As mentioned above, most large-scale savings in investments are obtained by reducing the need for stationary batteries. Figure \ref{fig:Instbat} summarizes the accumulated impact on storage capacity investment in the model over the whole period.
In the PC scenario, the model invests massively in grid-scale battery technologies in order to limit grid expansion. The investment in battery storage reaches 11,874 $M\EUR$ in $PC_{noTransEx}$ whereas transmission expansion can reduce expenditure by 1.3\%. The introduction of smart-charging schemes consistently limits the need for stationary batteries in reducing investments by approximately 62\% and 67\% without and with investments in transmission respectively. Investments in stationary batteries are reduced by 80\% to 93\% when V2G charging is associated with EV under the scenario of no transmission grid expansion and transmission grid expansion respectively.\\

\subsubsection{Electricity generation throughout the decades}
\begin{figure*}[h!]
	\centering
    \subfloat[Differences in energy production of the scenarios without transmission expansion with $PC_{noTransEx}$ as the base scenario.]{\includegraphics[width=0.8\columnwidth]			           {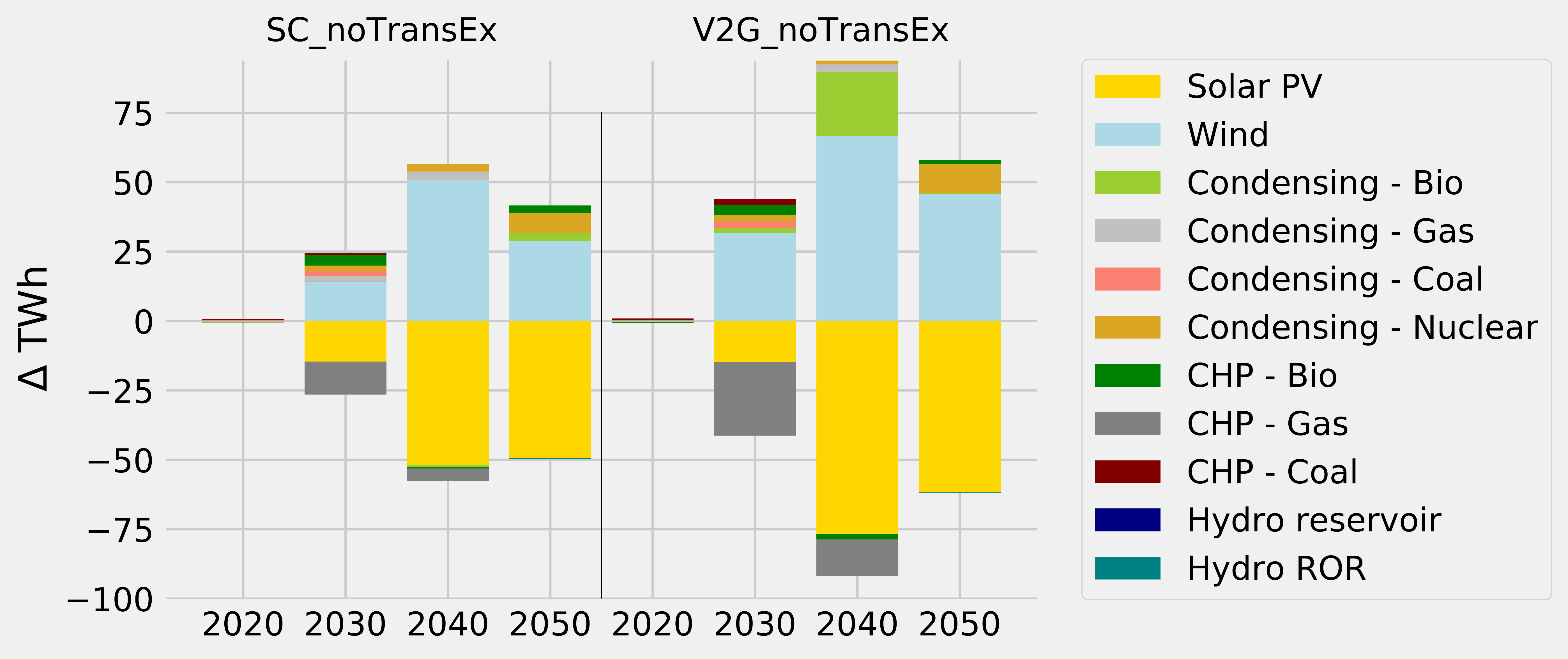}} \\
    \subfloat[Differences in energy production of the scenarios with transmission expansion with $PC_{TransEx}$ as the base scenario. ]{\includegraphics[width=0.8\columnwidth]
   	{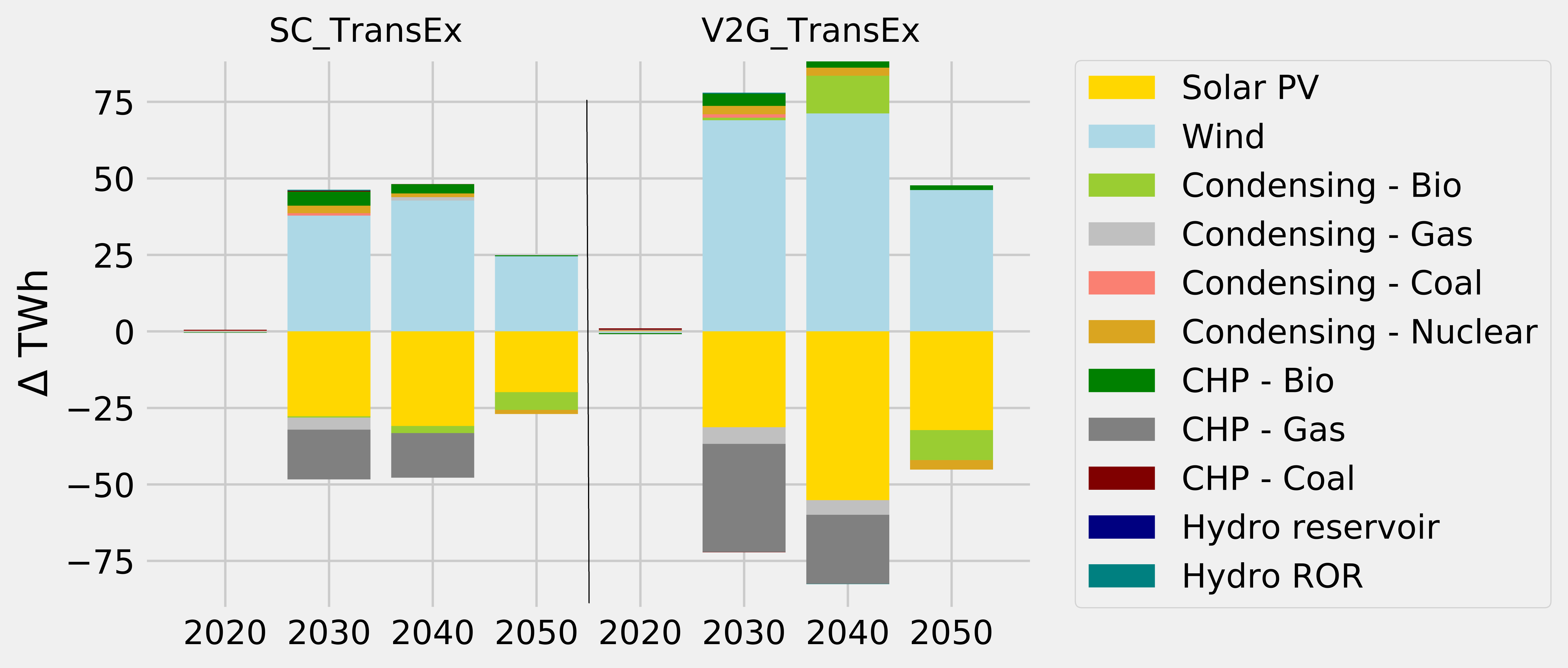}} \\
\centering
\caption{Absolute differences in electricity production by technology for the overall region of $SC_{noTransEx}$ and $V2G_{noTransEx}$ over the years by technology.}
\label{fig:ElGEnDiff}
\captionsetup[subfigure]{labelformat=empty}    
\end{figure*}

The different charging schemes also offer flexibility that substitutes to other technologies in the electricity and heat mix.
Figure \ref{fig:ElGenCap} shows the evolution in the generation of electricity in the base case, $PC_{noTransEx}$. 
\begin{figure}[H] 
    \centering
    \includegraphics[width=0.8\columnwidth]{Figures/El_production_PC_noTransEx.png} 
    \caption{Composition of electricity production in $PC_{noTransEx}$ over the years} 
    \label{fig:ElGenCap} 
\end{figure} 
The largest shift in generating technologies, regardless of the EV charging scenario, takes place in the decade from 2020 to 2030. Energy production by wind turbines and solar panels becomes dominant in the system, representing 52\% of electricity production as early as 2030. By contrast, coal in CHP and condensing units is phased out quickly, whereas natural gas still plays a role until 2040, with 3\% of the electricity supply share. The share of nuclear energy decreases in the 2020s due to the phasing out of ageing power plants and then remains relatively stable and substantial until 2050, especially due to investments in France and Finland. Altogether, wind, solar and nuclear energy represent 66\% of total electricity supply in 2050, or 1019 $TWh$, 473 $TWh$, and  430 $TWh$ respectively. CHP and condensing power plants burning biofuels contribute marginally to energy production ($<$6.2\% in year 2050). The model shows a more intense usage of biofuels in the heating sector. However, with the ongoing transition in the transport sector, biofuel prices will increase more dramatically, which will subsequently reduce the competitiveness and business opportunities of this resource in the heat sector.\\
The base case for transmission expansion $PC_{TransEx}$ develops similar towards 2050. A figure \ref{fig:PassiveTEELprod} can be found in the appendix \ref{sec;elprodpc}. The greatest differences are to be found in the early years with plants using gas. Due to stronger market coupling, less peak capacity is needed to cover peak demand because additional transmission lines can serve additional flexibility in $PC_{TransEx}$. This effect amounts to 22.3\% and 30.3\% less gas consumption in the entire modelling region for 2030 and 2040. Solar PV provides approximately 382 $TWh$ and Wind 1120 $TWh$ in 2050. This represents a decline for solar power 19.3\% and a rise for wind of 9.9\% compared to $PC_{noTransEx}$. Biofuels represent only 5.3\% of the total energy production in 2050, whereas nuclear provides around 27 $TWh$ more in contrast to $PC_{noTransEx}$ yielding an overall market share of approximately 19\%. The main effect on nuclear power is explained by stronger market coupling and greater flexibility provided by additional transmission lines, thus improving business cases for baseload technologies at the expense of technologies delivering peak power.
\\\\
Figure \ref{fig:ElGEnDiff} shows the differences in electricity production per year compared to $PC_{noTransEx}$ without transmission expansion in the upper figure and to $PC_{TransEx}$ with transmission expansion in the lower figure.
The results show that the increased flexibility of V2G, compared to SC and the base case, mainly benefits wind development and to some extent the participation of biomass-based plants. As soon as 2030, introducing $SC {noTransEx}$ and $V2G_{noTransEx}$ results in an increase in wind production of respectively 14 $TWh$  and 32 $TWh$  compared to the base case. In 2050, the two scenarios also support an increase in wind-power generation of up to 46 $TWh$. The difference is relatively smaller due to the growing size of the wind sector.\\

However, flexible charging schemes have an opposite effect on solar and gas-based CHP generation. PV participation decreases continuously compared to the base case, along with the growing flexibility generated by SC and V2G. The main difference in output appears in the 2040s, when the main investment in solar PV is made in the base case. $SC_{noTransEx}$ results in 50 $TWh$ of avoided solar production compared to $PC_{noTransEx}$. \\
Solar PV mainly produces during high price hours throughout the day. EVs contribute with load-shifting and sell energy which helps to integrate more wind power. Longer periods between low and high wind production hours can be bridged. At the same time, investments in the cheapest available technology option are encouraged, regardless of production pattern. \\
The main difference in production from gas-based CHP occurs during the 2030s, given the still substantial relative share of this technology in the mix (-27TWh in V2G). Like solar PV, the increasing flexibility provided by the charging schemes limits the need to run gas turbines. Nevertheless, condensing power plants burning biofuels remain the cheapest resource to contribute to system balance at peak hours. At the same time, baseload power such as nuclear power increases its output with a load-shifting capacity of up to 11 $TWh$ and commits to the replacement of peak power capacity.\\

All the effects depicted above are further intensified with transmission grid expansion and better market coupling. In particular, the scenarios with transmission investments further limit the participation of gas CHP during the 2030s, yielding to a reduction of 35 $TWh$ with V2G charging compared to the $PC_{TransEx}$ scenario. More transmission lines also support the greater integration of wind energy. The production of wind energy increases by 24-37 $TWh$ in $SC_{TransEx}$ and $V2G_{TransEx}$ compared to no expansion in $SC_{noTransEx}$ an $V2G_{noTransEx}$. 
In 2050, the additional flexibility of SC and V2G competes with heat pumps in the heat sector. Thus, around 15-20 $TWh$ less electricity is produced due to the greater level of competition. Power-to-heat technologies are mainly replaced by biomass-fuelled boilers, while the substitution effects are also cross-affected by the better use of transmission lines in $V2G_{noTransEx}$ charging strategies. With the additional flexibility obtained through grid expansion, the business case for heat pumps is not as strong as before under competition with EV affecting electricity prices. While in $V2G_{noTransEx}$ electricity production of condensing power plants using biofuels is stable compared to the other charging schemes, the case is certainly different with transmission lines. $V2G_{TransEx}$ schedules around 10 $TWh$ less than $V2G_{noTransEx}$ in 2050. A similar case is valid for $SC_{TransEx}$ with around -6 $TWh$.
The main exception is solar PV, where more network capacity results in less degradation of solar participation compared to both charging scenarios without grid investment. Ultimately, more interconnection damages the production from condensing gas plants, previously unaffected by the level of charging flexibility and the production of condensing biomass plants in the last decade. The reason for both these effects lies in the increased market coupling. More interconnections allow the use of cheaper wind capacities and the smoothing out of fluctuations on a spatial scale. At the same time, baseload production from nuclear stays constant with different charging schemes when expanding the grid. Finally, it is worth stressing that EV flexibility in particular affects technologies with variable outputs and costly peak covering technologies. The output of hydroelectricity, which remains competitive in order to provide flexibility, is not impacted by any of the charging schemes.

\subsubsection{Electricity prices fluctuations}
\begin{figure*}[h!]
	\centering
    \subfloat[Standard deviation of scenarios without transmission expansion. ]{\includegraphics[width=0.8\columnwidth]			           {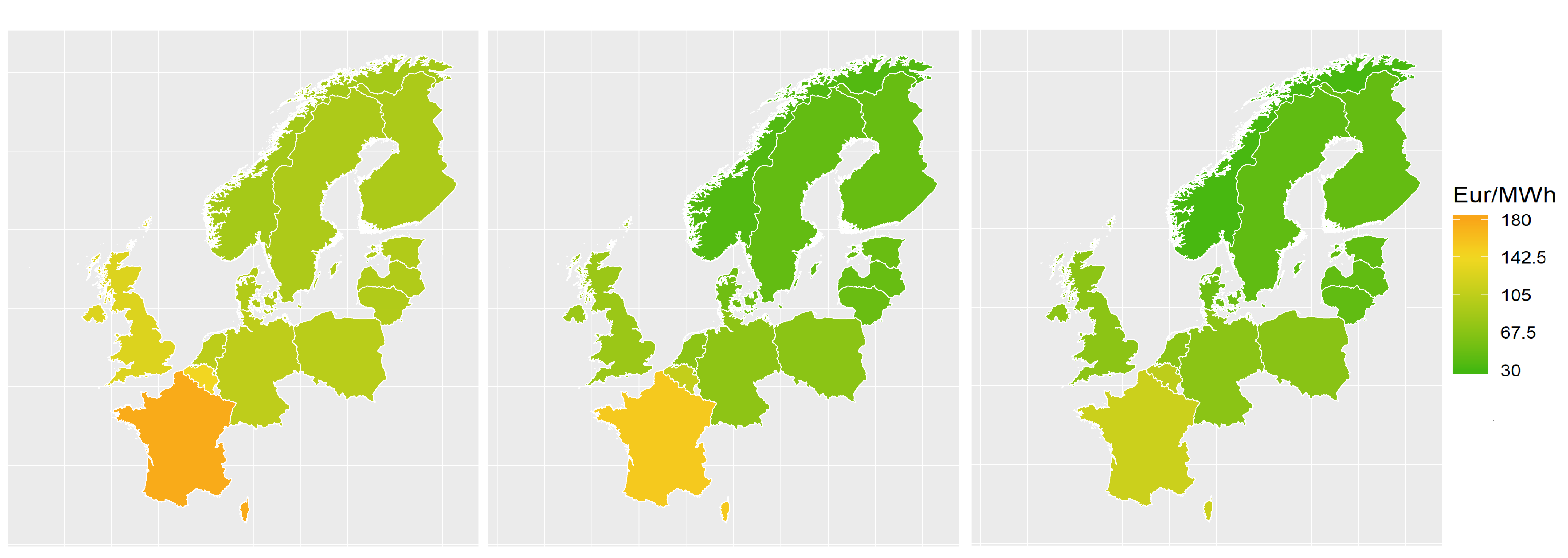}} \\
    \subfloat[Standard deviation of scenarios with transmission expansion. ]{\includegraphics[width=0.8\columnwidth]
   	{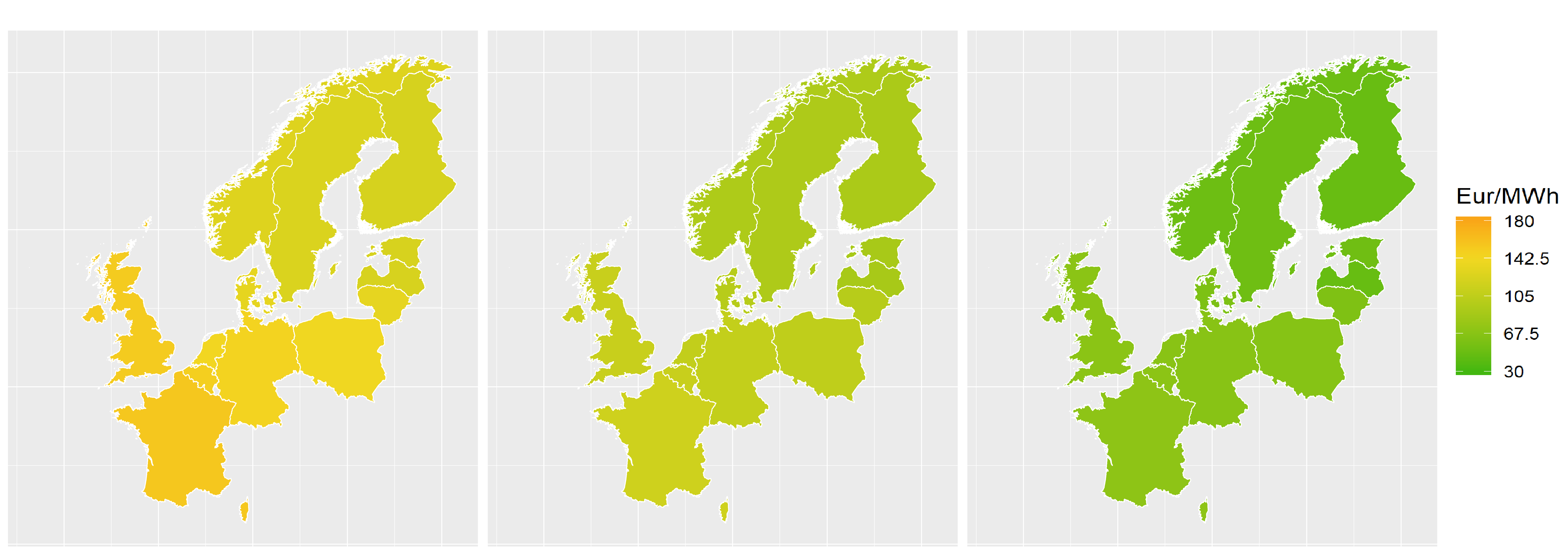}} \\
\centering
\caption{Standard deviation of electricity prices in $\EUR$/MWh with PC, SC and V2G from left to right in 2050.}
\label{fig:ElPriMod}
\captionsetup[subfigure]{labelformat=empty}    
\end{figure*}

Figure \ref{fig:ElPriMod} illustrates the change in the standard deviation of prices when increasing the flexibility of EVs with and without transmission system expansion in 2050.
As Balmorel runs in foresight mode, it can see incoming vehicles from the following year and thus optimises the system based on the anticipated demand increase. Thus, the charging schemes influence investment decisions and electricity prices. Consequently, the applied charging pattern already influences investment decisions in generating and transmission capacities in earlier years and amplifies the effects until 2050. The way the different charging schemes affect electricity prices is therefore twofold. Flexible EV charging significantly smooths out spot price variations, but at the expense of a marginally higher average spot price.  \\

The highest standard deviation in the modelling region is located in France for $PC_{noTransEx}$ with 176 $\EUR$/MWh. This due to a large share of nuclear power production with a limited ramp capacity. However, passively charging vehicles require rapidly responding power plants during peak hours. Surrounding countries such as Belgium are further affected by price fluctuations due to interconnection. The lowest standard deviation of prices is located in Scandinavia, in particular in Norway. Highly flexible hydropower buffers the incoming fleet of inflexible EVs. Consequently, the price variations are limited to a standard deviation of 87 $\EUR$/MWh. Introducing smart charging already reduces variation significantly. In $SC_{noTransEx}$ France reduces the standard deviation of prices by 13\%, whereas Norway achieves a reduction of as much as 57\% yielding in just 37$\EUR$/MWh in variation. Contrasting $PC_{noTransEx}$ with $V2G_{noTransEx}$ further shows an improvement for price stability. France now achieves a standard deviation of 113 $\EUR$/MWh corresponding to 36\% whereas Norway now has a reduction of 63\%. Countries dominated by inflexible generating capacity in particular improve price stability by using the full flexibility of EVs with V2G. Compared to countries with production flexibility such as Norway, smart charging already reduces price variation the most. \\

Similar tendencies in the direction of price variations can be seen in the three lower cases with transmission expansion. At first, prices are smoothed out throughout the modelling region. Electricity prices show a standard deviation in France for the $PC_{TransEx}$ scenarios of 155 $\EUR$/MWh, which represents a 12\% reduction compared to $PC_{noTransEx}$. However, most of the other countries therefore face higher standard deviations in order to reduce the fluctuations in the south-west of the modelling region. Norway, for example, now has a  greater standard deviation in electricity prices for $PC_{TransEx}$ compared to no transmission expansion in $PC_{noTransEx}$. This strong effect is due to the smaller production capacity of Norway compared to France and the optimality of a holistic price stability. Smart charging improves these outcomes significantly over the entire modelling region. The lowest price variation in $SC_{TransEx}$ is 93 $\EUR$/MWh whereas the largest is only 116 $\EUR$/MWh in France. $V2G_{TransEx}$ achieves the most stable outcome with regards to price fluctuations. France yields into 71 $\EUR$/MWh while Norway hovers around 51 $\EUR$/MWh. The differences on a spatial scale here are also relatively small compared to $V2G_{noTransEx}$ with a maximum spread of 82 $\EUR$/MWh.\\

Since the objective of the model is to reduce total system costs for all regions, it attempts to reduce the fluctuations, in particular in the south-western countries. Reducing the price variations in large consumption centers like France, the UK and Germany has a stronger positive effect on reducing the system costs. Consequently, favorable low emission potentials in Scandinavia are utilized more by larger investments into renewable technologies. The increased production is subsequently shared via larger interconnection capacities with neighboring countries. This however also exposes countries like Norway to larger price fluctuations, leading to electricity price convergence. As Scandinavia overall has less market share, its rise in price fluctuations is larger than the reduction in the rest of the modelling region.\\ 

With fewer fluctuations, the weighted average prices only increase slightly. When loads are shifted, the natural effect is a reduction in electricity prices during peak hours and an increase in prices during hours with previously lower prices. This effect is observed system-wide. The weighted average electricity prices only increase between 2.3-3.5\% with smart charging and V2G compared to $PC_{noTransEx}$ with 42.45 $\EUR$/MWh as the base case in 2050. $PC_{TransEx}$ has a weighted average electricity price over the entire modelling region of 39.85 $\EUR$/MWh, whereas $SC_{noTransEx}$ and $V2G_{noTransEx}$ only yield into 39.86 $\EUR$/MWh and 39.89 $\EUR$/MWh respectively. \\
All in all, electricity price fluctuations are smoothed out the greater the degree of flexibility provided by EV. Smart charging has effects especially on countries with little flexibility on the generating side, whereas V2G supports highly flexible systems in particular. Transmission expansion reduces the spatial differences in electricity price fluctuations. The reduction in fluctuations is replaced by a slight increase in average prices due to EV load-shifting.

\subsubsection{Reduction of CO2 emissions from flexibility provision}

\begin{figure*}[h!]
	\centering
    \subfloat[Differences in CO2 cumulated CO2 emission in scenario without transmission expansion compared to $PC_{noTransEx}$. ]{\includegraphics[width=0.8\columnwidth]			           {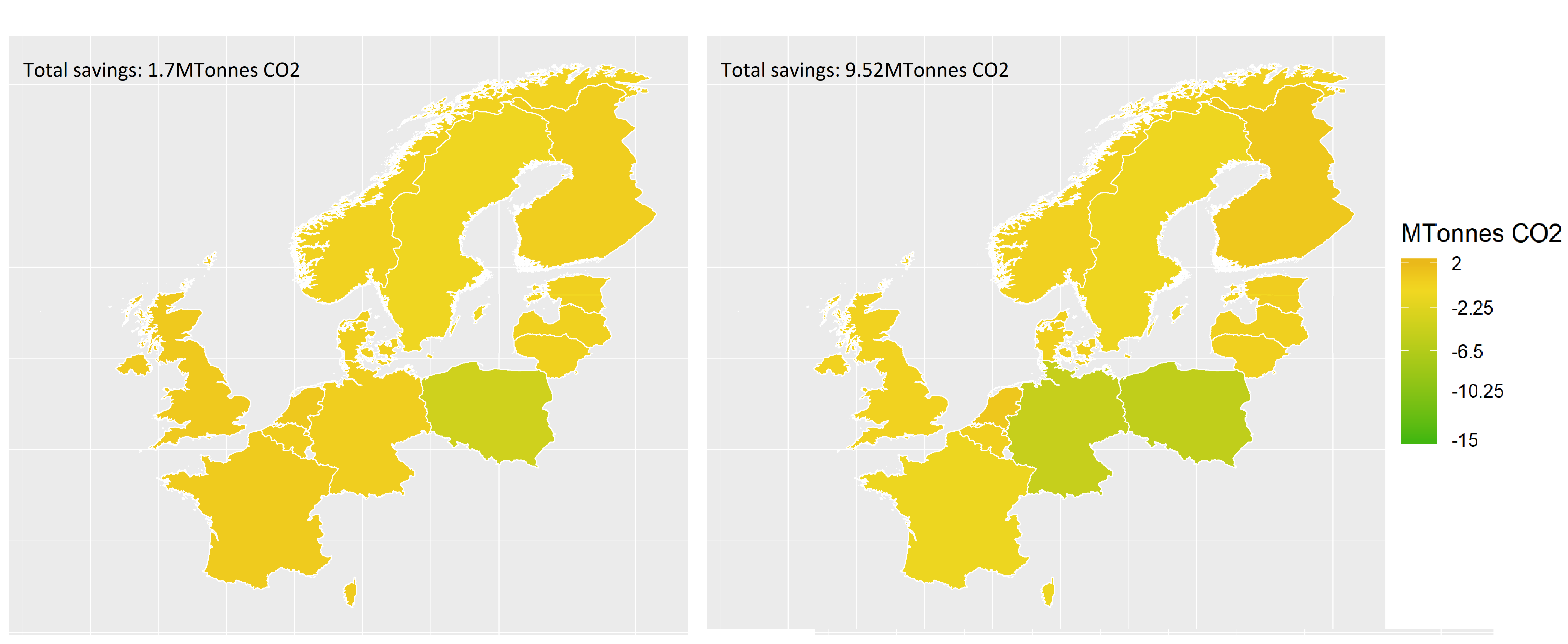}} \\
    \subfloat[Differences in CO2 cumulated CO2 emission in scenario without transmission expansion compared to $PC_{TransEx}$. ]{\includegraphics[width=0.8\columnwidth]
   	{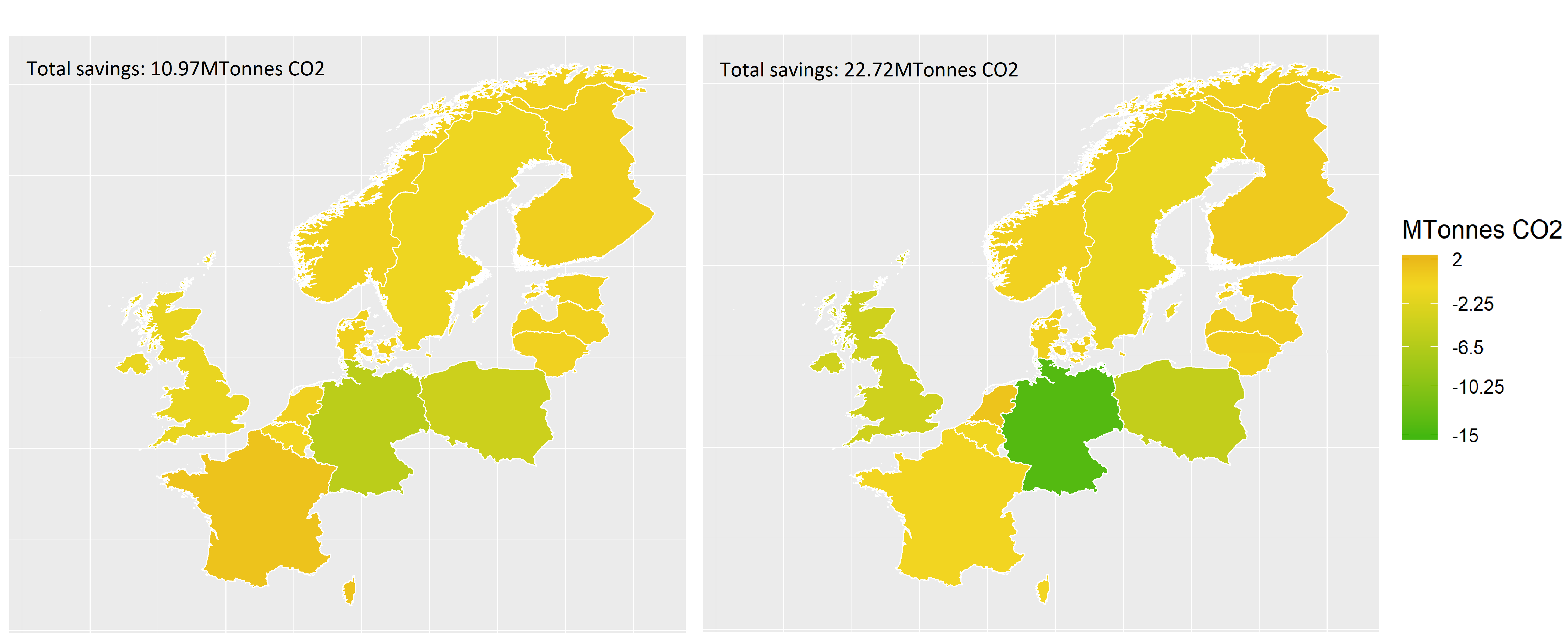}} \\
\centering
\caption{Total accumulated difference of CO2 emissions in $MTonnes$ until 2050 per country compared to their respective PC base case.}
\label{fig:EMI}
\captionsetup[subfigure]{labelformat=empty}    
\end{figure*}

The impact of V2G on reductions in CO2 emissions is shown in Figure \ref{fig:EMI}. The map summarizes the accumulated changes generated by a switch from PC to SC and V2G in all scenarios.
Our results show a significant system-wide reduction in CO2 emissions. The total savings over the period in $SC_{noTransEx}$ amount to 1.7 $MTonnes$ CO2 and to 9.5 $MTonnes$ CO2 with $V2G_{noTransEx}$, which corresponds to a reduction of approximately 1.2\% compared to the base case. While Germany can barely save GHG in $SC_{noTransEx}$, V2G has positive effects of approximately 4.6 $MTonnes$. Poland is the largest positive contributor with 3.8 and 5.23 $MTonnes$. What are striking are the distributional effects, since some countries end up emitting more CO2, in particular Denmark and the Netherlands. Given continued fossil fuel-based energy production in Poland and Germany, the flexible charging schemes not only help integrate more VRE into the system, they also improve the utilization of the grid in supporting lower emission technologies with higher efficiencies, namely CHP in Denmark and gas power plants in the Netherlands, allowing energy production from coal to be replaced.\\
Given additional transmission expansion $SC_{TransEx}$ the emission savings were increased by factor of 6.5 to 11 $MTonnes$. A reduction of 239\% is achieved with V2G between with and without grid investments, which amounts to 22.8 $MTonnes$. By solving bottlenecks in Germany, $SC_{TransEx}$ already results in 5.7 $MTonnes$ fewer emissions than in $PC_{TransEx}$, while Poland approximately maintains its previous levels at 4.2 $MTonnes$. Transmission expansion and V2G have by far the greatest effects on Germany, contributing to a reduction of 14 $MTonnes$, but the UK too now reduces emissions considerably with 3.8 $MTonnes$. Conversely, Denmark and the Netherlands increase their emissions slightly, including compared to the $_{noTrans}$, by approximately 23\%. However, due to the low absolute values, the overall avoided CO2 emissions in the entire region offsets the growth in the small countries.

To summarize the result section, the following bullet points highlight the most relevant outcomes of the modelling exercise.

\begin{itemize}
  \item Both EV flexibility through smart charging and V2G as well as transmission expansion reduce system costs considerably.
  \item While flexibility provided by transmission expansion only had limited effects on investments into stationary storage, smart charging and V2G had the largest impact.
  \item Due to the progressive increase in CO2 taxes, the electricity system will achieve carbon neutrality in 2050, with energy production from wind, solar and nuclear power, hydropower and biomass in descending order.
  \item Changing EV charging schemes results in substitution effects from solar PV in combination with batteries towards more wind production from passive charging towards V2G.
  \item Transmission expansion after 2030 intensifies the substitution of solar PV by cheaper wind resources and limits the usage of peak power production units such as gas plants.
  \item While passive charging schemes result in increased electrification of the heat sector, flexible charging of EVs limits the business case for heat pumps to operate, resulting in reduced deployment of power-to-heat technologies.
  \item EV flexibility has strong local effects on the reduction of electricity price fluctuations in contrast to transmission flexibility, which smoothens price fluctuations on spatial scale.
  \item Flexibility of EV and transmission expansion accelerates CO2 reduction in CO2 intensive countries, but at the expense of the carbon footprint of more virtuous countries. 
  \item Regardless of the distributional effect in terms of CO2 emissions, the overall regional positive environmental effect stresses that strong European coordination is needed to achieve deep decarbonisation target in 2050.

\end{itemize}

\section{Discussion} \label{sec:Disc}

In summary, the results presented in section \ref{sec:Res} demonstrate that the flexibility enabled by EV charging schemes generates efficiency gains across the entire energy system, which translates into lower costs, higher VRE penetration and lower CO2 emissions.\\
The progressive rise in CO2 prices is the main driver for RES investments, as it forces the model to shift early into decarbonized energy sources. Thus the integration of RES, especially wind and biomass energy, is already substantial and only marginally improved by flexible EV charging. Nevertheless, the results stress the strong combination effect that occurs by linking flexible EVs to wind penetration, which reveals a twofold dynamics. First, a virtuous effect is established between the two technologies where the greater flexibility accelerates the penetration of wind energy, further improving the competitiveness of electricity as a fuel for cars. Second, a substitution effect arises between flexible EV charging and solar penetration. Solar PV produces mainly during high price hours throughout the day. EVs contribute with their load-shifting flexibility and by selling energy to integrate more wind power. As a consequence, longer periods between low and high wind production hours can be bridged. At the same time, this promotes investments in the cheapest available technology option, regardless of the production pattern. Our results underline the substitution effect between solar and wind induced by EV but also transmission flexibility. This substitution effect arises through underlying assumptions as well as the modelling approach, which have not been modelled in detail in large scale energy system models before. In response, solar power production is reduced since consumption coverage is not necessarily needed that much anymore during the day. Another, simultaneous explanation is the given pattern of availability. As charging at work is not an option in the present study and as vehicles are not available during those hours, the model recognizes less need for solar. This effect further supports wind penetration. By comparing the outcomes to Brown et al. \cite{Brown2018SynergiesSystem} the substitution effect is expected to be less significant with greater vehicle availability during the day, and special use cases like commercial vehicles are not taken into account \cite{Andersen2018AddedServices}. With charging at work, it is expected to have a positive effect on solar PV as the EV can utilize the production pattern better. At the same time, peak power plants such as gas power plants reduce their production with increasing flexible EV charging. This is also an effect of load-shifting reducing peaks and with them the profitable hours for peak technologies. \\
A fixed transmission system converts the shifted consumption on to increased baseload production. The effect is less severe when transmission expansion offers flexibility on the spatial level in order to be balanced with more VRE technologies. The corresponding effect of only EV flexibility supporting baseload capacities was also recognized by Hedegaard \cite{Hedegaard2013WindModelling}. Fixed transmission further supports the deployment of solar PV. \cite{Gils2017IntegratedEurope} and \cite{Horsch2017TheScenarios} have shown similar effects between the role of transmission expansion plans and investments into solar power. Also \cite{Brown2018SynergiesSystem} has shown model sensitivities in terms of wind integration dependent on transmission system development.\\
In addition, our results underline the key substitution effects that exist between competing technologies. Flexible EV charging associated with a critical mass of electric vehicles also questions the competitiveness of stationary batteries. It is, however, worth noting that this study does not consider the provision of ancillary services from either electric car batteries or stationary batteries. Including the additional flexibility and revenues associated with balancing services might change this conclusion. At the same time, \cite{Brown2018SynergiesSystem} presents similar reduction of investments into stationary batteries with rising EV flexibility. The substitution effects are not limited to the electricity sector, as EV flexibility also has knock-on effects on the business opportunities for P2H in district heating, as it competes directly with heat pumps especially, but also with electric boilers. P2H is replaced by biomass-fuelled boilers and is further intensified by more interconnections. However, given the ongoing transition of the transport sector, including aviation and on-road freight transport, biofuel prices are expected to increase more dramatically than what the model represents, which will subsequently reduce the competitiveness and business opportunities of biofuels in the heating sector. In this case, the business case of P2H may be restored and improved. \\
The penetration of electric cars is not expected to have a dramatic impact on the level of wholesale electricity prices, but the lower price volatility allowed by flexible EV charging clearly reflects the better use of cheap resources in time and, through a ripple effect, of lower CO2 emissions. Focusing at the country level shows substantial distribution effects in CO2 reductions as further shown in \cite{Flex4RES2019FlexibleReport}. In particular, better flexibility and the market coupling negatively affect Denmark and the Netherlands, as both countries support the transition of more carbon-intensive countries in respectively operating their gas-based CHPs and condensing power plants more as replacements for German and Polish coal plants, and delaying their phasing out. This emphasizes that EU countries should cooperate throughout the transition period, rather than focus on domestic-only targets if they want to achieve the greatest impact at the European level.\\

Larger transmission investments in V2G are likely due to the spatial granularity of the model’s cost assumptions for grid expansion. A rough spatial granularity reduces the visibility of bottlenecks and the complexity of power flow optimization. Therefore, the model sees fewer complex problems to increase single line capacities which is also the case  in \cite{Horsch2017TheScenarios} and \cite{Gunkel2020ModellingStudy}. With more detailed transmission system representation, adjacent lines are also affected. Consequently, more lines have to be extended, increasing the costs. The merit order of different options for generating investments versus transmission expansion is thus decisive for that outcome. Additionally, transmission line projects have greater variation in prices, and consequently uncertainty hovers over these outcomes regarding whether they are realistic enough. At the same time, substitution effects from peak capacities towards more wind with transmission expansion are also seen in \cite{Chen2020TheEurope}. It is valid for the Scandinavian energy system but also central and western Europe. \cite{Chen2020TheEurope} further shows distributional effects on European scale. Additional grid expansion results into larger wind capacities in Northern Europe to utilize the favorable resources in order to support Central Europe in their decarbonization efforts. Similarly, our results show that Poland and Germany are the main beneficiaries in terms of CO2 reductions. \\

At the same time, regulatory framework barriers are hindering the utilization of EVs flexibility. In particular, restrictions in terms of market entry barriers and minimum bid size restrict profitable market participation strategies \cite{Kester2018PromotingDiffusion}. In particular, restrictions in terms of market entry barriers such as minimum bid size or not adapted technical standards limit EVs in offering their flexibility to the market \cite{Kester2018PromotingDiffusion}. It is important to provide sufficient incentives for business models and EV owners to use the flexibility of their vehicles to help the system. The impact of EVs on transmission levels may seem limited, but the greatest potentials likely when running a model allow for stacked services from low- to high-level, including grid services.\\

Two restrictive factors could limit the optimal response further. On the one hand, due to a more realistic risk-hedging bidding strategy the day before, smoother participation in the markets can be expected. On the other hand, the vehicles will further respond to congestion problems and voltage issues in the distribution level, which moreover changes the reaction to price signals in upper-level markets. Although these points are challenges to the flexible behavior of EVs on the transmission level, they also offer solutions for distribution grids by offering reactive power, as well as active congestion management, which might become even more critical \cite{Gjelaj2019GridDemand} \cite{Andersen2019TheV2G}.  Due to their complexity, distribution system costs are not included in the optimization. Consequently this study only provides potentials from the perspective of day-ahead markets and moreover disregards relevant potentials and revenue streams in balancing and frequency markets \cite{Gjelaj2019GridDemand}, \cite{Andersen2018AddedServices}.
Moreover, it is not likely that the whole fleet will be available for flexibility due to the human factor. Vehicles that are usually flexible may also be charged due to spontaneous intermediate SOC goals. Also, the temperature dependency of demand due to heating and AC systems is currently not being implemented. Moreover, bidirectional charger prices are still an uncertain factor and must be studied in greater depth, since they can affect the feasibility of V2G charging substantially, their costs being significant in the first decades.\\

\section{Conclusion}\label{sec:conc}
This paper has investigated the impact of electric vehicle charging schemes and transmission expansions on long-term energy-system planning by focusing on investment decisions in generating capacity and transmission lines, electricity prices, electricity and heat production, and CO2 emissions. The scenarios cover passive charging, smart charging and vehicle-to-grid, each with and without the option of transmission system expansion. \\\\
Increasing the flexibility of electric vehicles reduces system costs considerably and also suggests synergies with transmission expansion plans. At the same time, investments in stationary batteries are strongly affected by the introduction of flexible electric vehicles, whereas additional transmission expansion has only marginal effects. Moreover, smart charging and vehicle-to-grid compete with other sector-coupling technologies such as heat pumps and electric boilers. The focus on home charging in combination with the introduction of electric vehicle flexibility triggers the replacement of solar PV by wind generation capacities due to the load-shifting capability, which is not severely hampered by calendar and cycle degradation. Without transmission expansion, costly peak capacities like gas power plants are replaced with baseload. On the other hand, given endogenous transmission expansion, variable renewable energy is installed in larger quantities due to better market coupling and spatial smoothing. Flexible electric vehicle charging shows only marginal effects on average prices but reduces price variability significantly, which reflects the better use of energy resources and results in cuts in CO2 emissions.\\
However, the integration of flexible behavior faces significant regulatory barriers which need to be tackled, such as minimum bid requirements, aggregation and technical approval. Future studies may focus more on ancillary services to improve the business case for technologies serving flexibility, as well as taking operations in distribution systems into account. At the same time, the impact of charging outside the home and a general expansion of the transport sector should also be studied given limited availability of biofuels.

\section*{Acknowledgment}
Philipp Andreas Gunkel, Ida Græsted Jensen and Claire Bergaentzlé prepared this paper as part of the Flex4RES (www.flex4RES.org) research project, which is supported by Nordic Energy Research (http://www.nordicenergy.org/), under the auspices of Nordic Council of Ministers, Norway [contractnumber:76084]. The authors appreciate the opportunity to work on this topic and would like to send special thanks to the project partners, organizations and reviewers. Fabian Scheller receives funding from the European Union's Horizon 2020 research and innovation programme under the Marie Sklodowska-Curie grant agreement no. 713683 (COFUNDfellowsDTU). The usual disclaimer applies.

\nomenclature[A,01]{$PC$}{Passive charging} 
\nomenclature[A,02]{$SC$}{Smart charging} 
\nomenclature[A,03]{$V2G$}{Vehicle-to-grid charging} 
\nomenclature[A,03]{$VRE$}{Variable renewable energy} 
\nomenclature[A,03]{$EV$}{Electric vehicle} 
\nomenclature[A,03]{$PHEV$}{Plug-in-hybrid electric vehicle} 
\nomenclature[A,03]{$SOC$}{State of charge} 
\nomenclature[A,03]{$CWE$}{Central West Europe} 
\nomenclature[A,03]{$DOD$}{Depth of discharge} 

\nomenclature[C,01]{$\Delta t$}{Length of a time step} 
\nomenclature[C,01]{$C^{Bat}$}{Battery investment cost [€]}
\nomenclature[C,01]{$C^{Ch}$}{Charger investment cost [€]}
\nomenclature[C,02]{$\gamma^{Bat,Cyc}$}{Battery cycle factor} 
\nomenclature[C,02]{$\gamma^{Cal,Const}$}{Unavoidable battery aging factor} 
\nomenclature[C,02]{$\gamma^{Cal,Flex}$}{Flexible battery aging factor} 
\nomenclature[C,03]{$\alpha^{Bat,Os}$}{Battery availability factor} 
\nomenclature[C,03]{$\alpha^{Bat,Lft}$}{Battery lifetime factor} 
\nomenclature[C,03]{$\alpha^{Deg,CalC}$}{Battery fixed aging factor} 
\nomenclature[C,03]{$\alpha^{Deg,CalF}$}{Battery varibale aging factor} 
\nomenclature[C,03]{$\eta^{Bat,Ch}$}{Battery charging efficiencies}
\nomenclature[C,01]{$\overline{P}^{Ch}$}{Charger size [MW]}
\nomenclature[C,01]{$\overline{SOC}^{PHEV}$}{Battery storage size of PHEV [MWh]}
\nomenclature[C,01]{$\overline{SOC}^{BEV}$}{Battery storage size of BEV [MWh]}

\nomenclature[I,01]{$y$}{Year}
\nomenclature[I,02]{$a$}{Area}
\nomenclature[I,03]{$v$}{Vehicle}
\nomenclature[I,04]{$s$}{Season}
\nomenclature[I,05]{$t$}{Hour}

\nomenclature[P,01]{$P_{y,a,v,s,t}^{PC}$}{Predefined PC load [MW]}

\nomenclature[P,01]{$P_{y,a,v,s,t}^{Inflex}$}{Inflexible SC or V2G load [MW]}
\nomenclature[P,01]{$C_{y,a,v,s,t}^{Bat, Repl}$}{Battery replacement cost [€]}
\nomenclature[P,01]{${SOC}_{y,a,v,s,t}^{PC}$}{Battery state of charge PC [MWh]}

\nomenclature[P,01]{$\overline{SOC}_{y,a,v,s,t}$}{Maximum battery capacity of the fleet [MWh]}
\nomenclature[P,01]{$\underline{SOC}_{y,a,v,s,t}$}{Minimum battery capacity of the fleet [MWh]}
\nomenclature[P,01]{$Tr_{y,a,v,s,t}$}{Vehicle trips energy consumption [MWh]}
\nomenclature[P,01]{$\overline{P}_{y,v}^{Ch}$}{Maximum power of charger [MW]}
\nomenclature[P,01]{$\underline{P}_{y,v}^{Ch}$}{Minimum power of charger [MW]}
\nomenclature[C,03]{$Qty_{c,s,t}^{Bat,Avail}$}{Amount of available vehicles}

\nomenclature[V,01]{$\Omega_{y,s,t}$}{Charging demand [MWh]}
\nomenclature[V,01]{$\Phi_{y,s,t}^{Deg,Cyc}$}{Battery cycle degradiation cost [€]}
\nomenclature[V,01]{$\Phi_{y,s,t}^{PC,Deg,Cal}$}{Battery calendar degradiation cost [€]}
\nomenclature[V,01]{$VSOC_{y,a,v,s,t}$}{Battery state of charge [MWh]}
\nomenclature[P,01]{$VP_{y,a,v,s,t}^{Flex}$}{Flexible SC or V2G charging load [MW]}
\nomenclature[P,01]{$VP_{y,a,v,s,t}^{Flex}$}{Flexible V2G discharging load [MW]}

\printnomenclature

\bibliographystyle{elsarticle-num}
\bibliography{references}

\appendix

\section{Appendix}
\subsection{Algorithm to obtain vehicle availability} \label{sec;alg}

\begin{algorithm}[H]
\scriptsize
\caption{Generation of fleet characteristics}\label{alg:VehiclePattern}
\begin{algorithmic}[]
\State Set $c\in$$\mathcal{C}$, $t\in$$\mathcal{T}$, $d\in$$\mathcal{D}$
     \For{$d$ in $\mathcal{D}$}{}
         \For{$c$ in $\mathcal{C}$}{}
            \State Random pick from $\Omega (d)$$\Rightarrow \omega_{d,c} (t^{Dep},t^{Arr},l^{Dist})$\Comment{Random pick}
             \For{t in $\mathcal{T}$}{}\Comment{Availability}
                 \If{$t < t_{d,c}^{Dep}$ or $t > t_{d,c}^{Arr}$}{}
                     \State $Qty_{d,c,t}^{Avail}=1$
                 \EndIf
                 \If{$t \geq t_{d,c}^{Dep}$ or $t \leq t_{d,c}^{Arr}$}{}
                     \State $Qty_{d,c,t}^{Avail}=0$
                \EndIf
             \EndFor
             \For{t in $\mathcal{T}$}{}\Comment{Flexible charging}
                 \If{$t = t_{d,c}^{Dep}$}{}
                     \State $Tr_{d,c,t}^{Flex}=\overline{SOC}^{v}$
                 \EndIf
                 \If{$t \neq t_{d,c}^{Dep}$}{}
                     \State $Tr_{d,c,t}^{Flex}=0$
                 \EndIf
             \EndFor
             \For{t in $\mathcal{T}$}{}\Comment{Inflexible charging, flexibility after trip and dumb demand}
                  \If{$t = t_{d,c}^{Arr}+1$}{}
                     \If{$(l_{d,c,t}^{Dist} \cdot cons^{v}) > 0$}{}
                        \If{$(l_{d,c,t}^{Dist} \cdot cons^{v}) \leq \overline{P}^{Ch}$}{}
                            \State $P_{d,c,t}^{PC}=l_{d,c,t}^{Dist}\cdot cons^{v}$
                            \State $SOC_{d,c,t}^{PC}=\overline{SOC}^{v}$
                        \EndIf 
                        \If{$(l_{d,c,t}^{Dist} \cdot cons^{v}) > \overline{P}^{Ch}$}{}
                            \State $SOC_{d,c,t}^{PC}=\overline{SOC}^{v}-(l_{d,c,t}^{Dist} \cdot cons^{v})+\overline{P}^{Ch}$
                            \State $P_{d,c,t}^{PC}=\overline{P}^{Ch}$
                        \EndIf
                     \EndIf
                     \If{$SOC^{Em} < \overline{SOC}^{v}-(l_{d,c,t}^{Dist} \cdot cons^{v})$}{}
                         \State $P_{d,c,t}^{Inflex}=SOC^{Em}-(\overline{SOC}^{v}-(l_{d,c,t}^{Dist}\cdot cons^{v})) $
                         \State $SOC_{d,c,t}^{Flex}=\overline{SOC}^{v}$
                     \EndIf
                     \If{$t = t_{d,c}^{Arr}+1$}{}
                         \State $P_{d,c,t}^{Inflex}=0$
                         \State $SOC_{d,c,t}^{Flex}=\overline{SOC}^{v}-(l_{d,c,t}^{Dist}\cdot cons^{v})$
                     \EndIf
                 \EndIf
                 \If{$t \neq t_{d,c}^{Arr}+1$}{}
                     \State $P_{d,c,t}^{Inflex}=0$
                     \State $SOC_{d,c,t}^{Flex}=0$
                     \If{$\overline{SOC}^{v}-SOC_{d,c,t-1}^{PC} \leq \overline{P}^{Ch}$}{}
                        \State $P_{d,c,t}^{PC}=\overline{SOC}^{v}-SOC_{d,c,t-1}^{PC}$
                     \EndIf 
                     \If{$\overline{SOC}^{v}-SOC_{d,c,t-1}^{PC} > \overline{P}^{Ch}$}{}
                        \State $SOC_{d,c,t}^{PC}=SOC_{d,c,t-1}^{PC}+\overline{P}^{Ch}$
                        \State $P_{d,c,t}^{PC}=\overline{P}^{Ch}$
                     \EndIf
                 \EndIf
             \EndFor
             \For{t in $\mathcal{T}$}{}\Comment{Max and min SOC}
                 \State $\overline{SOC}_{d,c,t}=Qty_{d,c,t}^{Avail} \cdot \overline{SOC}^{v}$
                 \State $\underline{SOC}_{d,c,t}=Qty_{d,c,t}^{Avail} \cdot SOC^{Em}$
             \EndFor
         \EndFor
     \State $Qty_{c,s,t}^{Bat,Avail}=\sum_{c=0}^{\mathcal{C}}Qty_{d,c,t}^{Avail}$
     \State $P_{d,t}^{Flex}=\sum_{c=0}^{\mathcal{C}}Tr_{d,c,t}^{Flex}$
     \State $P_{d,t}^{Inflex}=\sum_{c=0}^{\mathcal{C}}P_{d,c,t}^{Inflex}$
     \State $P_{d,t}^{PC}=\sum_{c=0}^{\mathcal{C}}P_{d,c,t}^{PC}$
     \State $SOC_{d,t}^{Flex}=\sum_{c=0}^{\mathcal{C}}SOC_{d,c,t}^{Flex}$
     \State $\overline{SOC}_{d,t}=\sum_{c=0}^{\mathcal{C}}\overline{SOC}_{d,c,t}$
     \State $\underline{SOC}_{d,t}=\sum_{c=0}^{\mathcal{C}}\underline{SOC}_{d,c,t}$
     \State $SOC_{d,t}^{PC}=\sum_{c=0}^{\mathcal{C}}SOC_{d,c,t}^{PC}$
     \EndFor   
     \State \textbf{return} $Qty_{c,s,t}^{Bat,Avail},P_{d,t}^{Flex},P_{d,t}^{Inflex},P_{d,t}^{PC},SOC_{d,t}^{Flex},\overline{SOC}_{d,t},\underline{SOC}_{d,t}$
\end{algorithmic}
\end{algorithm}

\subsection{Electricity production in $PC_{TransEx}$} \label{sec;elprodpc}
\begin{figure}[H] 
    \centering
    \includegraphics[width=0.8\columnwidth]{Figures/El_production_PC_TransEx.png} 
    \caption{Composition of electricity production in $PC_{TransEx}$ over the years} 
    \label{fig:PassiveTEELprod} 
\end{figure}


\end{document}